\documentclass[
reprint,
aps,
floatfix,
]{revtex4-2}
\usepackage[utf8]{inputenc}
\usepackage{amsmath}
\usepackage{amssymb}
\usepackage{stackrel}
\usepackage{babel}
\usepackage{dcolumn}
\usepackage{bm}
\usepackage{graphicx} 
\usepackage{pgf} 
\usepackage{placeins}
\usepackage{lipsum}

\begin{document}
\preprint{APS/123-QED}
\title{A mathematical theory for understanding when abstract representations emerge in neural networks}


\author{Bin Wang} 
\email{bw2841@columbia.edu.}
\author{W. Jeffrey Johnston}
\author{Stefano Fusi}%
 \email{sf2237@columbia.edu.}
\affiliation{%
 Center for Theoretical Neuroscience, Columbia University, New York, NY
}%

\date{\today}



\begin{abstract}
Recent experiments in neuroscience reveal that task-relevant variables are often encoded in approximately orthogonal subspaces of neural population activity. These disentangled, or abstract, representations have been observed in multiple brain areas and across different species. These representations have been shown to support out-of-distribution generalization and rapid learning of novel tasks. The mechanisms by which these representations emerge remain poorly understood, especially in the case of supervised task behavior. Here, we show mathematically that abstract representations of latent variables are guaranteed to appear in the hidden layer of feedforward nonlinear networks when they are trained on tasks that depend directly on these latent variables. These learned abstract representations reflect the semantics of the input stimuli. 
To show this, we reformulate the usual optimization over the network weights into a mean‐field optimization problem over the distribution of neural preactivations. We then apply this framework to finite-width ReLU networks and show that the hidden layer of these networks will exhibit an abstract representation at all global minima of the task objective. Finally, we extend our findings to two broad families of activation functions as well as deep feedforward architectures. 
Together, our results provide an explanation for the widely observed abstract representations in both the brain and artificial neural networks. In addition, the general framework that we develop here provides a mathematically tractable toolkit for understanding the emergence of different kinds of representations in task-optimized, feature-learning network models.
\end{abstract}

\maketitle

\section{Introduction}
How task structure shapes the geometry of neural representations has long been a central question in neuroscience and machine learning. The ability to learn appropriate representations from training data is fundamental to a system's capacity for generalization. In neuroscience, recent experiments have shown that, following task training, neural responses across various brain regions often exhibit a characteristic low-dimensional geometry, referred to as an abstract (or, in machine learning, a disentangled) representation \cite{bernardi2020geometry,courellis2024abstract,nogueira2023geometry,boyle2024tuned,tang2023geometric,fascianelli2024neural,o2023representational,mishchanchuk2024hidden} (Fig.~\ref{fig:1}). 
In these representations, task-relevant variables are represented in distinct, approximately orthogonal subspaces in the firing rate space of neurons (Fig.~\ref{fig:1}A, right). 
One possibility is that these subspaces correspond to distinct sets of neurons. In this case, a single neuron would represent the value of a single variable. The discovery of an unsupervised learning algorithm that will produce this kind of "disentangled" representation has been a topic of great interest in machine learning \cite{higgins2017beta,mathieu2019disentangling,fernandez2023disentangling}. Another possibility is that these subspaces are not necessarily aligned with single neurons, but instead distributed across the activity of the entire population of neurons, as seen in many experimental work from neuroscience \cite{bernardi2020geometry,courellis2024abstract,fascianelli2024neural}, where all neurons in the population play at least a small part in representing almost all variables. Whether expressed in specialized sub-populations or across the whole population, abstract representations make the representation of a single variable invariant to all other variables ('dissociated from specific instances'). As a consequence, they facilitate certain forms of out-of-distribution generalization and rapid learning of novel tasks. In contrast, non-abstract representations (Fig.~\ref{fig:1}B, right) -- in which different variables are not invariant to each other -- do not have these properties, though the relevant variables can still be accurately decoded \cite{fusi2016neurons,kaufman2022implications}. In these non-abstract representations, the different variables are nonlinearly mixed with each other \cite{rigotti2013importance,fusi2016neurons} -- producing a high-dimensional representation that can be used to decode any stimulus condition dichotomy. 

In machine learning, variational autoencoders and their variants have been the standard approach to obtain disentangled representations \cite{burgess2018understanding}. However, due to identifiability issues, it has been shown that learning disentangled representations in an entirely unsupervised manner is difficult if not impossible \cite{hyvarinen1999nonlinear,locatello2019challenging,locatello2020sober}. For this reason, other approaches have been proposed, which often include additional regularization \cite{horan2021unsupervised,arjovsky2017towards,fernandez2023disentangling,liang2025compositional} or supervision \cite{ruder2017overview,johnston2023abstract,alleman2024task}. These supervised approaches have a clear connection to neuroscience: through natural behavior, humans and other animals learn to interact with and classify stimuli in their environment in a variety of ways. Previous work has shown that learning to classify single stimuli in multiple ways can naturally give rise to abstract representations in artificial neural networks \cite{johnston2023abstract}. Thus, abstract representations may emerge in the brain through this behavioral process. However, this prior work did not establish general conditions in which these representations emerge, nor did they develop a mathematical theory that explains this emergence. 

Here, we formulate an analytical theory that explains the emergence of abstract representations in feedforward artificial neural networks trained to perform multiple tasks that share the same set of latent variables \cite{johnston2023abstract,alleman2024task}. We begin by studying the emergence of abstract representations in the simplest nonlinear network model that exhibits representation learning: a feedforward network with a single hidden layer and nonlinear activation functions (Fig.~\ref{fig:1b}A, left). For the first time, we prove that the process of minimization of a simple mean square error with $l^2$-weight regularization in the multi-task setting of \cite{johnston2023abstract,alleman2024task} is guaranteed to generate abstract representations. We show that this result is robust to the choice of nonlinear activation function. However, our analytical framework can be applied well beyond the tasks studied here. In particular, it can be used to characterize the optimal neural representation in nonlinear feature-learning neural networks trained on any task. Our framework provides a powerful tool for characterizing the structure of neural representation in task-optimized networks \cite{richards2019deep,yang2020artificial,saxe2021if,kanwisher2023using,doerig2023neuroconnectionist}. Our work therefore advances existing analytical methods for studying task-optimized neural networks, and more broadly, in those nonlinear models exhibiting permutation symmetry.

The paper is organized as follows. In Section~\ref{sec:The-analytical-framework}, we define the task and network model and then introduce the analytical framework for characterizing the optimal neural representations in a two-layer nonlinear network trained on a given task. This framework establishes an exact mapping from the original network model to an effective model whose degrees of freedom are the neural preactivation patterns on training data (Fig.~\ref{fig:1c}A). Solving for the optimal neural representation is then reduced to a corresponding mean-field theory of this effective model. In Section~\ref{sec:Whitened-and-target-aligned}, we apply this framework to finite-width ReLU networks and derive the optimal neural representation for the corresponding task model (Fig.~\ref{fig:1c}B). In Section \ref{sec:Abstract-representation-emerges}, we further use this analytical framework to show that the abstract representation remains optimal for two broad classes of nonlinear activation functions. Finally, in Section~\ref{sec:Extensions-and-other}, we extend the framework to study additional tasks and deep network architectures. 

\section{The analytical framework\label{sec:The-analytical-framework}}
\subsection{Data and network model \label{subsec:Data-and-network}}

We consider feedforward network models trained through supervised learning, with the training dataset
\begin{equation}\label{Eq:DataModel}
\mathcal{D}=\{\left(\mathbf{x}^{i},\mathbf{y}^{i}\right)\}_{i=1}^{P}\subseteq\mathbb{R}^{d_{X}}\times\mathbb{R}^{d_{Y}}.
\end{equation}
Here, $P$ is the number of training samples (input-output mappings) in the task. $d_X$ and $d_Y$ are the input and output dimensions.

\begin{figure}[tbp]
    \centering
    \includegraphics[width=\linewidth]{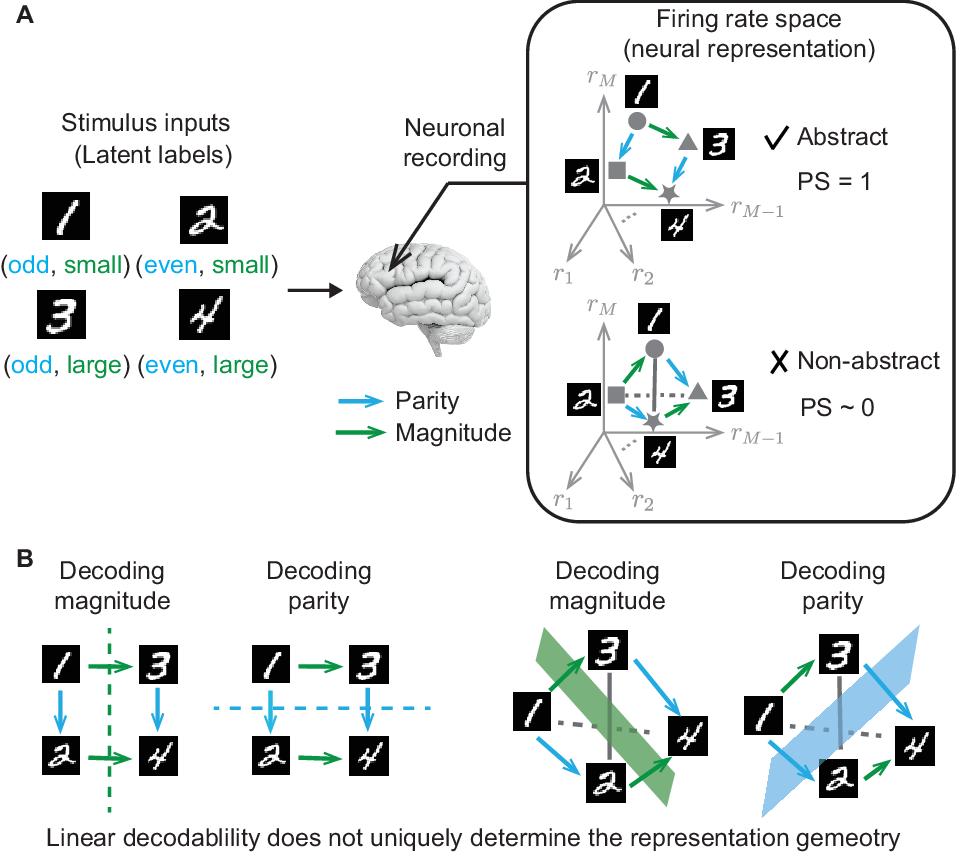}
    \caption{Abstract representations. (A) Each stimulus input in the task (e.g. images of handwritten digits) is associated with several binary latent variables (e.g. parity and magnitude of the digits). These latent variables are exactly the output labels $\mathbf{y}^i$ in our data model Eq.~\eqref{Eq:DataModel}. An abstract representation is one in which each binary variable is represented along a single axis in the population activity space, shown in the top plot inside the frame. This geometric property can be quantified using the parallelism score ($PS$), which measures how parallel the coding directions for one variable remain when the other variables vary. For example, it would measure the parallelism of the parity coding direction for small and large digits (corresponding to two values of the other variable, magnitude). Abstract representations are low-dimensional and have $PS=1$. In an alternative neural representation, the points representing the different digits are arranged on a tetrahedron shape, which is the highest-dimensional representation (bottom in the frame). This would correspond to a non-abstract representation and has $PS\sim0$. (B) Magnitude and parity are equally linearly decodable for both geometries.}
    \label{fig:1}
\end{figure}


We assume that the input patterns are unstructured (that is, uncorrelated with each other), whereas the desired outputs (i.e., the training labels) have low-dimensional structure (similar to \cite{saxe2019mathematical,johnston2023abstract}).
In general, these labels might reflect the structure of a latent space used to generate the complex, seemingly unstructured inputs and might depend on the latent variables in a complex way \cite{johnston2023abstract}. Here, we assume that the network's desired outputs, $\mathbf{y}^i$, are exactly these latent variables, and ignore the possibility of a more complex dependency between the latent variables and output labels. An example of such dataset is the MNIST images where the latent variables are magnitude and parity (Fig.~\ref{fig:1}A).


More specifically, the output associated with a particular input $\mathbf{y}^{i} \in \{\pm 1\}^{d_Y}$ is assumed to consist of $d_{Y}\in\mathbb{N}_{+}$ binary labels. In the typical multitask set-up \cite{ruder2017overview}, different binary labels represent different ways to classify the inputs, usually based on different features of the inputs (Fig.~\ref{fig:1}). We define the input and output data matrices as
\begin{align}\label{eq:input_output_mat}
X_{data}&=\left(\mathbf{x}^{1},\mathbf{x}^{2},...,\mathbf{x}^{P}\right)\in\mathbb{R}^{d_{X}\times P}, \nonumber \\ Y&=\left(\mathbf{y}^{1},\mathbf{y}^{2},...,\mathbf{\mathbf{y}}^{P}\right)\in\{\pm1\}^{d_{Y}\times P}.
\end{align}

Based on their output labels, all the training data form $2^{d_Y}$ distinct classes. We further assume that each class has the same number of training data, $n\in\mathbb{N}_{+}$. Namely, all the classes are balanced, and thus the total number of training data points $P$ satisfies 
\begin{equation}
P=n\cdot2^{d_{Y}}.\label{eq:P_2dy}
\end{equation}

Denote the $i$th row vector of $Y$ as $\mathbf{v}_{i}\in \{\pm 1\}^{P}$, whose components are the $i$th binary labels for all training data, $i=1,2,..d_Y$. Under the above assumption, we find that (i) $\mathbf{v}_i\cdot \mathbf{v}_j = P\delta_{ij}$, where $\delta_{ij}$ is the Kronecker delta notation; (ii) $\mathbf{1}\cdot\mathbf{v}_{i}=0$, for $\mathbf{1}=(1,1,...,1)^T\in\mathbb{\mathbb{R}}^{P}$, i.e. exactly half of the components of $\mathbf{v}_i$ is $+1$ and the other half is $-1$, and (iii) $\mathbf{v}_i$'s are all the eigenvectors of the output kernel matrix $K_Y \equiv Y^TY$ with non-vanishing eigenvalues (SI \S 1.1).

We are interested in the neural representation in a minimal two-layer network model trained to produce these input-output mappings (Fig.~\ref{fig:1b}). A two-layer network defines a map 
\begin{equation*}
f_{W_1,W_2,\mathbf{b}}(\mathbf{x})=W_2\phi(W_1\mathbf{x}+\mathbf{b}),
\end{equation*}
where $\phi$ is a component-wise nonlinear activation function. The width of the hidden layer is $M$. $W_1\in\mathbb{R}^{M\times d_{X}}$ and $W_2\in\mathbb{R}^{d_{Y}\times M}$ are the weight matrices of the 1st and 2nd layer. $\mathbf{b}\in\mathbb{R}^{M}$ is the bias parameter. When $M$ is large, the above functional form can approximate any continuous function \cite{cybenko1989approximation}. 

The weight and bias parameters in the network are optimized on the training dataset $\mathcal{D}$ via the loss function 
\begin{align} \label{eq:TwoLayerNN_loss}
E&(W_1,W_2,\mathbf{b}) \nonumber \\
=&\sum_{i=1}^{P}\left[\mathbf{y}^{i}-W_2\phi(W_1\mathbf{x}^{i}+\mathbf{b})\right]^{2}
 \nonumber \\
&\hspace{3.7em}+\lambda_{1}\left\Vert W_1\right\Vert _{F}^{2}+\lambda_{1}\left\Vert \mathbf{b}\right\Vert _{2}^{2}+\lambda_{2}\left\Vert W_2\right\Vert _{F}^{2}\nonumber \\
\equiv&\left\Vert Y-W_2\phi(WX)\right\Vert^{2}_F+\lambda_{1}\left\Vert W\right\Vert _{F}^{2} +\lambda_{2}\left\Vert W_2\right\Vert _{F}^{2}. 
\end{align}
Here $\left\Vert \cdot\right\Vert _{F}$ is the Frobenius norm of a matrix. In the last line, we introduced the augmented input matrix $X$ and weight matrix $W$ to incorporate the bias parameter $\mathbf{b}$,
\begin{equation}\label{eq:aug_matrix}
W\equiv\left(W_1,\mathbf{b}\right),\quad X\equiv\left(\begin{array}{c}
X_{data}\\
\mathbf{1}^{T}
\end{array}\right).
\end{equation}

$\lambda_{1,2}>0$ in Eq.~\eqref{eq:TwoLayerNN_loss} are the strengths of $l^{2}$-weight regularization and are typically small. 

For a data point $(\mathbf{x},\mathbf{y})$, we call $\mathbf{r}=\phi(W_1\mathbf{x}+\mathbf{b})$ the population $\textit{firing rate vector}$ or $\textit{neural representation}$ of the data in the hidden layer. We assume that after the task training, the neural network has reached a global minimum of the loss [Eq.~\eqref{eq:TwoLayerNN_loss}]. And we will investigate the neural representations at these global minima. Our main analysis and results will be presented in this setting of two-layer neural networks, and the extensions to deep neural networks are deferred to Section \ref{sec:Extensions-and-other}.

\subsection{Parallelism score measures the abstractness of the neural representation
\label{subsec:Parallelism-score-measures}}

Previous work uses \textit{parallelism score} (PS) as a measure of the abstractness of a neural representation \cite{bernardi2020geometry,alleman2024task}. We define PS below using our notation. Given a network with the parameters $(W_1,W_2,b)$, its representation of the training data $(\mathbf{x}^{i},\mathbf{y}^{i})$ is $\mathbf{r}^{i}=\phi(W_1\mathbf{x}^{i}+\mathbf{b})$. Since all the training data form $2^{d_Y}$ distinct classes based on their output labels, we define the prototype representation of each class as the mean firing rate vector of the $n$ data points [Eq.\eqref{eq:P_2dy}] belonging to this class,
\begin{equation}
\mathbf{r}(\mathbf{y})=\frac{1}{n}\sum_{i:\mathbf{y}^{i}=\mathbf{y}}\mathbf{r}^{i}.\label{eq:prototype_rep}
\end{equation}
where $\mathbf{y} = (y_1,..,y_{d_Y})\in \{\pm 1\}^{d_Y}$ is the class label.

An abstract representation is characterized by the property that each latent binary label is encoded along a specific direction in the population firing rate space, independently of the other latent labels. To measure this, we examine how the neural representation changes when varying only the $k$th latent label $y_{k}$ while keeping other labels $\mathbf{y}_{\setminus k}=\alpha\in\{\pm1\}^{d_{Y}-1}$ fixed,
\begin{equation*}
\Delta\mathbf{r}(k;\alpha)=\mathbf{r}(y_{k}=+1,\mathbf{y}_{\setminus k}=\alpha)-\mathbf{r}(y_{k}=-1,\mathbf{y}_{\setminus k}=\alpha).
\end{equation*}
For different labels $\alpha_{1},\alpha_{2}\in\{\pm1\}^{d_{Y}-1}$, we quantify how consistent the direction of representation changes are when varying the $k$th label, using cosine similarity
\begin{equation*}
PS_{k}(\alpha_{1},\alpha_{2})\equiv\frac{\Delta\mathbf{r}(k;\alpha_{1})\cdot\Delta\mathbf{r}(k;\alpha_{2})}{\left\Vert \Delta\mathbf{r}(k;\alpha_{1})\right\Vert _{2}\left\Vert \Delta\mathbf{r}(k;\alpha_{2})\right\Vert _{2}} \in [-1,1].
\end{equation*}
From this definition, if the representation change for the $k$th latent label is independent of other latent labels $\left[\Delta\mathbf{r}(k;\alpha_{1})=\Delta\mathbf{r}(k;\alpha_{2})\right]$, then $PS_{k}(\alpha_{1},\alpha_{2})=1$. So any deviation from $1$ would indicate an interdependence between different latent labels. 

The average parallelism score for the $k$th latent label is defined as the average cosine similarity for all distinct pairs $\left(\alpha_{1},\alpha_{2}\right),$
\begin{equation*}
PS_{k}=\frac{1}{2^{d_{Y}-1}\cdot(2^{d_{Y}-1}-1)}\sum_{\substack{\alpha_{1},\alpha_{2}\in\{\pm1\}^{d_{Y}-1}\\
\alpha_{1}\neq\alpha_{2}
}
}PS_{k}(\alpha_{1},\alpha_{2}).
\end{equation*}
The overall $PS$ for the neural representation is the average over all latent labels
\begin{equation}\label{eq:PS}
PS=\frac{1}{d_{Y}}\sum_{k=1}^{d_{Y}}PS_{k}.
\end{equation}.

By definition, $PS$ of a neural representation only depends on the inner product between the neural representation of all data pairs, $\mathbf{r}^{i}\cdot\mathbf{r}^{j}$. So we define the representation kernel matrix $K\in\mathbb{R}^{P\times P}$ whose element is 
\begin{equation}\label{eq:Kij}
K_{ij}\equiv\mathbf{r}^{i}\cdot\mathbf{r}^{j} = \phi(W_1\mathbf{x}^{i}+\mathbf{b}) \cdot \phi(W_1\mathbf{x}^{j}+\mathbf{b})
\end{equation}
This kernel matrix fully determines the $PS$ for a neural representation. Moreover, following the definition, $PS$ does not change when (1) adding a constant to all elements of $K$, or (2) multiplying $K$ by a positive scalar. 

If the neural representation of data is completely random, $PS\sim 0$. A neural representation is called an abstract representation if the $PS$ is large and close to $1$ \cite{bernardi2020geometry,courellis2024abstract}. However, even a representation with $PS\sim 0$ allows for a linear decoder to recover the latent labels (Fig.~\ref{fig:1}A). Thus, the emergence of representations with high $PS$ is not a direct consequence of high task performance.

\subsection{The effective mean-field energy \label{subsec:The-effective-mean-field}}

We investigate the global minima of the loss function for the two-layer network models $E(W_1,W_2,\mathbf{b})$ [Eq.~\eqref{eq:TwoLayerNN_loss}]. These minima can also be thought of as the ground states of a physical system with a Hamiltonian $E(W_1,W_2,\mathbf{b})$. The corresponding Gibbs measure with inverse-temperature parameter $\beta$ is
\begin{equation}
p(W_1,W_2,\mathbf{b})=Z_{\beta}^{-1}\exp\left[-\beta E(W_1,W_2,\mathbf{b})\right],
\end{equation}
where  $Z_{\beta}\equiv \int_{W_1,W_2,\mathbf{b}}\exp\left[-\beta E(W_1,W_2,\mathbf{b})\right]dW_1dW_2d\mathbf{b}$ is the corresponding partition function and the free energy is $F_{\beta} \equiv \beta^{-1}\ln Z_{\beta}$. Specifically, the ground state of the system is related to the zero-temperature free energy $\lim_{\beta \rightarrow +\infty} F_{\beta}$.

\begin{figure}[tbp]
    \centering
    \includegraphics[width=\linewidth]{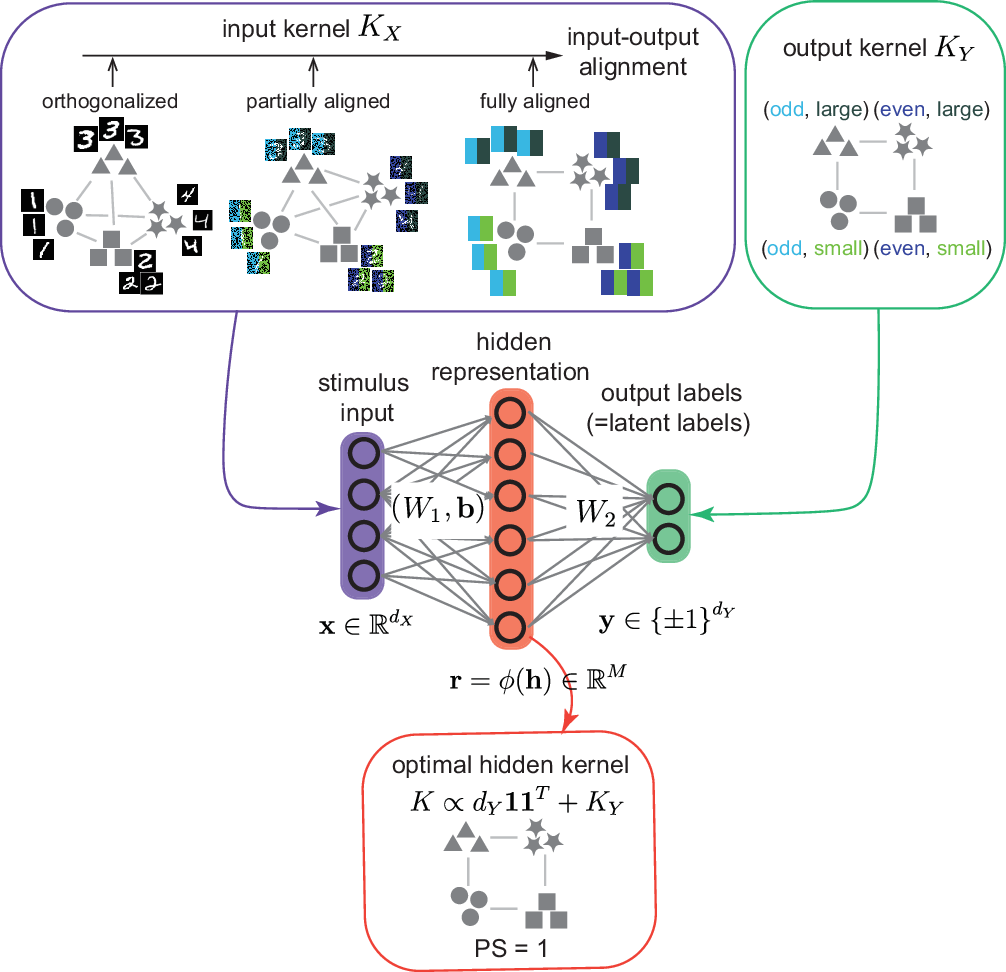}
    \caption{Model set-up and summary of the main results. (Middle) The two-layer nonlinear network models are trained on tasks related to abstract representation. The weight and bias parameters are optimized for the task. The hidden layer has width $M$. For a range of input geometries (characterized by different input kernel matrices $K_X$) and the specified output geometry (where each output label is exactly the latent label), the optimal hidden representation is abstract. (Top right) The output labels for each stimulus input are exactly its associated binary latent labels. (Top left) The range of input geometry smoothly transitions from a fully orthogonalized input, where different stimuli are represented by orthogonal vectors, to a fully output-aligned input (where $K_X \approx K_Y$). To illustrate the fully orthogonalized input in the $4$-dimensional space, we draw its 3D projection, which forms a tetrahedron.}
    \label{fig:1b}
\end{figure}

We introduce some notations to state our main result for the zero-temperature free energy. Denote the augmented input and output kernel matrices as 
\begin{equation}\label{eq:in_out_kernel}
    K_{X}=X^{T}X\in \mathbb{R}^{P\times P}, \quad K_{Y}=Y^{T}Y\in \mathbb{R}^{P\times P}.
\end{equation}
Since up to a rotation, the input and output data $X,Y$ can be reconstructed from these kernel matrices, we say that these kernel matrices capture the input and output {\em geometry} (Fig.~\ref{fig:1b}). Denote $\text{Range}K_X$ as the column space of the matrix $K_X$

The neural preactivations for all $P$ input data can be summarized in the preactivation matrix (Fig.~\ref{fig:1c}AB)
\begin{equation}\label{eq:preact_matrix}
    H \equiv X^TW^T = (\mathbf{h}_{1},\mathbf{h}_{2},...,\mathbf{h}_{M})\in \mathbb{R}^{P\times M},
\end{equation}
where the $k$th column vector $\mathbf{\mathbf{h}}_{k}\in\mathbb{R}^{P}$ represents the preactivation pattern of the $k$th hidden neuron for all $P$ data points (or stimulus conditions). We note that $H$ is related to the response matrix commonly used in neuroscience \cite{kaufman2022implications}, whose rows correspond to all the stimulus conditions and columns to all the neurons. 

We find that (SI \S 1.2) the zero-temperature free energy can be obtained via the following optimization problem over the neural preactivations,
\begin{equation}\label{eq:zero_temp_freeenergy}
    \lim_{\beta \rightarrow +\infty} F_{\beta} \quad = \underset{ \mathbf{h}_k\in \text{Range}K_X}{\text{min}}E(\mathbf{h}_1,\mathbf{h}_2,..,\mathbf{h}_M),
\end{equation}
where $E(\mathbf{h}_1,\mathbf{h}_2,..,\mathbf{h}_M)$ is
\begin{align}\label{eq:loss_h}
E&(\mathbf{h}_1,\mathbf{h}_2,..,\mathbf{h}_M) \nonumber \\
=&\lambda_{1}\sum_{k=1}^{M}\mathbf{h}_{k}^{T}K_{X}^{\dagger}\mathbf{h}_{k} +\text{tr}\left(\frac{\lambda_2}{\lambda_{2}+\sum_{k=1}^{M}\phi(\mathbf{h}_{k})\phi(\mathbf{h}_{k})^{T}}K_{Y}\right)
\end{align}
Here $K_{X}^{\dagger}$ is the Moore-Penrose inverse of $K_{X}$. Eqs.~\eqref{eq:zero_temp_freeenergy}-\eqref{eq:loss_h} shows that the zero-temperature free energy is determined by the global minima of the effective energy function [Eq.~\eqref{eq:loss_h}] over the neural preactivation patterns in the hidden layer, subject to the constraint $\mathbf{h}_k\in \text{Range}K_X$. 

This result maps the original system, whose energy function [Eq.~\eqref{eq:TwoLayerNN_loss}] is over its parameter space $(W_1,W_2,\mathbf{b})$, into an effective system with energy function described by Eq.~\eqref{eq:loss_h}. The effective system consists of $M$ neurons, each of which has a $P$-dimensional state variable $\mathbf{h}\in\mathbb{R}^{P}$, lying in the data-dependent subspace $\text{Range}K_{X}$ (Fig.~\ref{fig:1c}DE). Note that Eqs.~\eqref{eq:zero_temp_freeenergy}-\eqref{eq:loss_h} are valid for any network width $M$ and input/output kernel matrix $K_X/K_Y$. 

An immediate consequence of the above equation is that the optimal preactivation matrix $H_* = (\mathbf{h}_{1*},...,\mathbf{h}_{M*})$ [Eq.~\eqref{eq:preact_matrix}] and the corresponding optimal representation kernel $K_*=\phi(H_*)\phi(H_*)^T$ [Eq.~\eqref{eq:Kij}] only depend on the input and output kernel matrices $K_X$ and $K_Y$, rather than depending directly on the data matrices $X, Y$. This invariance is a general property for any rotationally-symmetric energy function such as Eq.~\eqref{eq:TwoLayerNN_loss}.

Eq.\eqref{eq:loss_h} is invariant under permutation over neurons (note that the $h_{k}$ only appear in summation). So we introduce the empirical measure of the preactivations $\rho_{M}=\sum_{k=1}^{M}\delta_{\mathbf{h}_{k}}$ (the unnormalized empirical distribution of $\mathbf{h}_k$'s) and rewrite the sums as integrals over this new distribution, so
\begin{align}\label{eq:loss_rho}
\int\lambda_1\mathbf{h}^{T}K_{X}^{\dagger}\mathbf{h}\,d\rho_{M}(\mathbf{h})
+\text{tr}\left(\frac{\lambda_{2}}{\lambda_{2}+\int\phi(\mathbf{h})\phi(\mathbf{h})^{T}\,d\rho_{M}(\mathbf{h})}K_{Y}\right) 
\end{align}
Note that the representation kernel can be written as $K[\rho_{M}]=\int\phi(\mathbf{h})\phi(\mathbf{h})^{T}\,d\rho_{M}(\mathbf{h})$. Since all the preactivation patterns $\mathbf{h}_k$'s lie in $\text{Range}K_{X}$, so is the support of the measure, $\text{supp}\rho_{M}\subseteq\text{Range}K_{X}$ 
\footnote{Here the support of a measure $\text{supp}\rho$ can be intuitively thought of as the largest region where the measure has nonzero mass, and it is formally defined as the largest closed set $A\subseteq\mathbb{R}^{P}$ such that $\rho(A^{c})=0$.}.

Eq.~\eqref{eq:loss_rho} shows that the energy function of the effective $M$-neuron system only depends on the global statistics of neural activity $\rho_{M}$. Thus, the measure $\rho_{M}$ can be viewed as the order parameter of the (finite-size) effective $M$-neuron system, which is permutation-invariant and reflects the symmetry in the original system [Eq.~\eqref{eq:TwoLayerNN_loss}]. This is similar to many other models in statistical physics \cite{parisi1980order,kuramoto1984chemical,kuramoto1975self,negele1982mean,jones2015density,opper2001advanced} where the order parameter here is a (probability) measure rather than simple scalars.

\subsection{The optimality condition}

To compute the $PS$ [Eq.~\eqref{eq:PS}] corresponding to the ground state of effective system, we want to find the representation kernel $K[\rho_{M}^{*}]$ corresponding to the global minima $\rho_{M}^{*}$ of Eq.~\eqref{eq:loss_rho}, with the constraint that $\rho_{M}^{*}$ is a sum of $M$ Dirac-delta measures and $\text{supp}\rho_{M}^{*}\subseteq\text{Range}K_{X}$. This optimization is challenging to perform directly, as the space of finite $M$-sum Dirac delta measures is not convex. To address this issue, we relax the constraint by enlarging the optimization domain to include all finite positive measures supported on $\text{Range}K_{X} \subseteq \mathbb{R}^P$, which also includes continuous measures. We denote this new space of measures \footnote{To be technically correct, we also require all 2nd-order moments of $\rho$ are finite in $M_{+}(X)$. This would ensure all the integrals in $E[\rho]$ are finite.}, as $M_{+}(K_X)$.
By definition, $M_+(K_X)$ is convex (equipped with the usual sum between measures).

\begin{figure*}[tbp]
    \centering
    \includegraphics[width=0.7\linewidth]{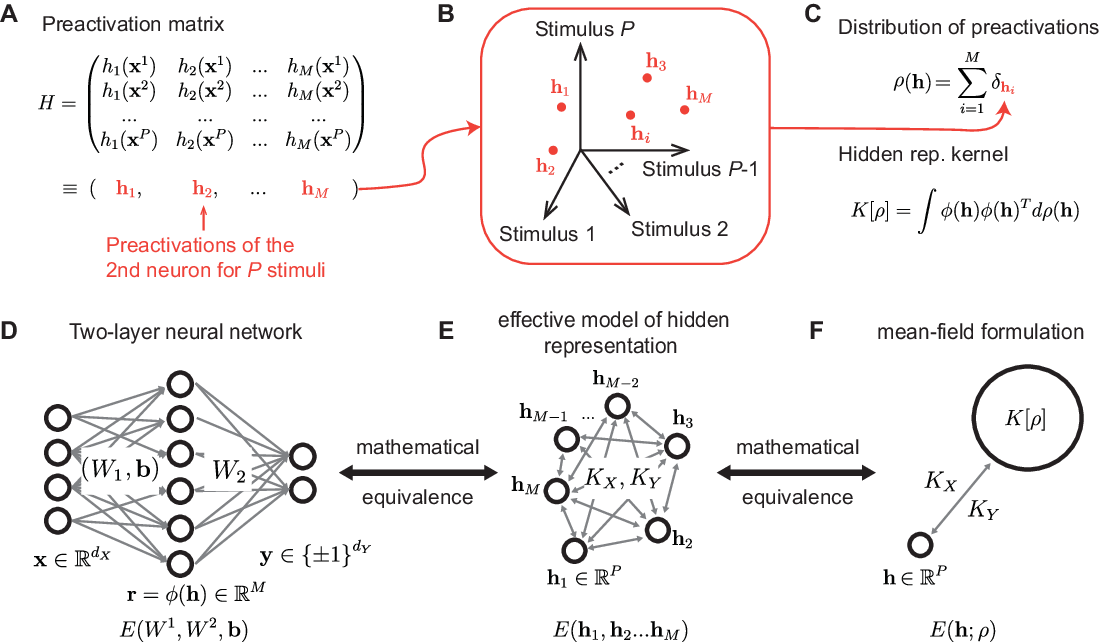}
    \caption{The analytical framework. (A) The neural preactivation patterns for all $P$ stimuli can be captured in the preactivation matrix $H$ [Eq.~\eqref{eq:preact_matrix}]. Each row represents the $M$ hidden neurons' preactivations for a specific stimulus. Each column represents the preactivations of a specific hidden neuron for all $P$ stimuli. (B) The column vectors of $H$ can be plotted in a $P$-dimensional space that encodes each hidden neuron's tuning for all $P$ stimuli. (C) The statistics of preactivations of hidden neurons can be captured by the empirical (unnormalized) measure. The hidden representation kernel matrix is a linear function of such an empirical measure. (D-F) Mathematically, finding the optimal network reduces to determining the ground state of an effective $M$-neuron system whose interactions are governed by the input and output kernel matrices [Eq.~\eqref{eq:loss_h}], which is further equivalent to a mean-field problem where a single representative neuron interacts with the statistics of the neural activity in the network [Eq.~\eqref{eq:MF_SingleNeuron}].}
    \label{fig:1c}
\end{figure*}

Now for $\rho \in M_{+}(X)$, we consider the energy functional defined by Eq.~\eqref{eq:loss_rho},  
\begin{equation}\label{eq:loss_functional}
E[\rho]\equiv\lambda_{1}\int\mathbf{h}^{T}K_{X}^{\dagger}\mathbf{h}\,d\rho(\mathbf{h})+\text{tr}\left[\frac{\lambda_{2}}{\lambda_{2}+\int\phi(\mathbf{h})\phi(\mathbf{h})^{T}\,d\rho(\mathbf{h})}K_{Y}\right].
\end{equation}

This is a convex functional on $M_{+}(X)$: $E[\rho]$ is a convex function on $M_{+}(X)$ if and only if $\forall\rho_{1},\rho_{2}\in M_{+}(X),\,f(t)\equiv E[t\rho_{1}+(1-t)\rho_{2}]$ is a convex function for $t\in[0,1]$ \cite{boyd2004convex}. The latter can be checked by computing the 2nd derivative of $f(t)$ (SI \S 1.3),
\begin{equation}\label{eq:loss_convexity}
f''(t)=2\lambda_{2}\text{tr}\left(\frac{1}{\lambda_{2}+K_{t}}\delta K\frac{1}{\lambda_{2}+K_{t}}K_{Y}\frac{1}{\lambda_{2}+K_{t}}\delta K\right)\geq0.
\end{equation}
See SI \S 1.3 for the expressions of $K_t$ and $\delta K$. 

For this convex optimization problem, the Karush--Kuhn--Tucker (KKT) condition for the optimal solution $\rho_{*}$ is (SI \S 1.4)
\begin{widetext}
\begin{align}\label{eq:KKT}
\lambda_{1}\mathbf{h}^{T}K_{X}^{\dagger}\mathbf{h}-\lambda_{2}\phi(\mathbf{h})^{T}\frac{1}{\lambda_{2}+K[\rho_{*}]}K_{Y}\frac{1}{\lambda_{2}+K[\rho_{*}]}\phi(\mathbf{h}) & \geq0,\forall\,\mathbf{h}\in\text{Range}K_{X},\nonumber \\
\mathbf{\,h}\in\text{supp}(\rho_{*})\, &\Rightarrow \,
\text{equality holds,}
\end{align}
\end{widetext}
where $K[\rho_{*}]=\int\phi(\mathbf{h})\phi(\mathbf{h})^{T}\,d\rho_{*}(\mathbf{h})$ is the representation kernel for $\rho_{*}$. Because the problem is convex in $\rho_{*}$, any solution to the KKT conditions is automatically a global minimum for  Eq.~\eqref{eq:loss_functional}. 



This KKT condition shows that any global minimum $\rho_{*}$ of Eq.~\eqref{eq:loss_functional} must satisfy the inequality in Eq.~\eqref{eq:KKT} and at the same time, its support must be confined to those vectors $\mathbf{h}$ for which the equal sign is attained. The two conditions can be interpreted as a mean-field description \cite{opper2001advanced} of the effective $M$-neuron system (Fig.~3F). To see this, we define the single-neuron mean-field energy,
\begin{align}\label{eq:MF_SingleNeuron}
E(\mathbf{h};&\rho)\equiv \lambda_{1}\mathbf{h}^{T}K_{X}^{\dagger}\mathbf{h}-\phi(\mathbf{h})^{T}\frac{\lambda_2}{\lambda_{2}+K[\rho]}K_{Y}\frac{1}{\lambda_{2}+K[\rho]}\phi(\mathbf{h}).
\end{align}
where $K[\rho]$ is the representation kernel for $\rho$, as defined above.

The KKT conditions [Eq.~\eqref{eq:KKT}] for any optimal solution $\rho_{*}$ can be written in terms of the single-neuron mean-field energy
\begin{align}\label{eq:SIngle_neuron_condition}
&\underset{\mathbf{h}\in\text{Range}K_{X}}{\text{min}}E(\mathbf{h};\rho_{*})=0, \nonumber \\
&\text{supp}(\rho_{*})\subseteq\underset{\mathbf{h}\in\text{Range}K_{X}}{\text{argmin}}E(\mathbf{h};\rho_{*}) \nonumber \\
&\quad \quad \quad \quad \equiv A(K_X,K_Y) ,
\end{align}
In the second equation, the support of $\rho_*$ consists of  the optimal preactivation patterns (or single-neuron tuning) for training data, which we denote as $A(K_X,K_Y)$. 

From statistical physics perspective, Eq.\eqref{eq:SIngle_neuron_condition} show that the individual neuron's preactivation $\mathbf{h}$ is trying to minimize the single-neuron mean-field energy $E(\mathbf{h};\rho)$, where the ``mean-field'' is generated by the statistics of all neurons' activity $\rho_{*}$ (Fig.~\ref{fig:1c}C). The two terms in $E(\mathbf{h};\rho)$ [Eq.\eqref{eq:MF_SingleNeuron}] have interesting interpretations: the 1st term pushes all the neurons' activities to align with the largest principal component of the input kernel, while the 2nd term encourages the transformed neuronal activity $\phi(\mathbf{h})$ to align with the largest principal component of the output-induced mean-field, $\frac{1}{\lambda_{2}+K[\rho_{*}]}K_{Y}\frac{1}{\lambda_{2}+K[\rho_{*}]}$. 

These mean-field equations [Eq.~\eqref{eq:SIngle_neuron_condition}] need to be solved in a self-consistent manner \cite{opper2001advanced,kuramoto1984chemical,kuramoto1975self,negele1982mean} (See SI \S 1.4). For arbitrary input and output data $K_X,K_Y$, and nonlinear activation function $\phi$, these equations are systems of nonlinear equations that can only be solved numerically. Fortunately, for training data related to abstract representation (Section \ref{subsec:Data-and-network}), the mean-field equations [Eq.~\eqref{eq:SIngle_neuron_condition}] can be solved exactly as we will present below. 


As a final note, although the above optimization is performed over all positive measures in $M_{+}(X)$ and in general, some of these measures cannot be attained by a finite-size system, in many cases, the global minimizer $\rho_{*}$ is a finite sum of Dirac-delta measures. As we show in the next section, this fact allows us to study the optimal neural representations in a {\em finite-width} network by optimizing a convex functional [Eq.~\eqref{eq:loss_functional}] over the infinite-dimensional measure space for preactivation, $M_{+}(X)$. 

\section{Whitened and target-aligned inputs lead to abstract representation in ReLU network \label{sec:Whitened-and-target-aligned}}


We investigate the optimal neural representations for the data and network model in Section \ref{subsec:Data-and-network}. Throughout this section, we assume a ReLU activation function: $\phi(z)=[z]_{+}$. We focus on two types of inputs: (i) whitened inputs, and (ii) inputs exhibiting stronger alignment with the outputs than the whitened case (hereafter “target-aligned inputs”, illustrated in Fig.~\ref{fig:1b}).  

For these input geometries, we find that all the solutions satisfying the KKT conditions [Eq.~\eqref{eq:SIngle_neuron_condition}] correspond to abstract hidden representations. We summarize the key steps of the argument here, with detailed derivations provided in SI \S 2. We start with the scenario where each binary class contains a single data point $(n=1)$ and then extend the results to multi-element classes $(n\geq2)$.

\subsection{Whitened input + single-element class ($n=1$) \label{subsec:Whitened-input-+}}

For whitened (or orthogonalized) inputs, $X_{data}^{T}X_{data}=I_{P}$. The augmented input kernel matrix and its psudo-inverse are
\begin{equation*}
K_{X}=I_{P}+\mathbf{1}\mathbf{1}^{T},\qquad K_{X}^{\dagger}=I_{P}-\frac{1}{P+1}\mathbf{1}\mathbf{1}^{T}.
\end{equation*}
Both matrices are full, so the constraint on $\mathbf{h}$ in Eq.~\eqref{eq:SIngle_neuron_condition} is trivial, $\text{Range}K_{X}=\mathbb{R}^{P}$. 

The key result that helps us solve the mean-field equation [Eq.\eqref{eq:SIngle_neuron_condition}] is the following lower bound for single-neuron mean-field energy (SI \S 2.1), 
\begin{align}\label{eq:whiten_1stterm}
    E(\mathbf{h};\rho) &\geq \lambda_{1}\mathbf{h}_+^{T}K_{X}^{\dagger}\mathbf{h}_+-\lambda_{2}\mathbf{h}_+^{T}\frac{1}{\lambda_{2}+K[\rho]}K_{Y}\frac{1}{\lambda_{2}+K[\rho]}\mathbf{h}_+  \nonumber \\
    &\equiv E_r(\mathbf{h}_{+};\rho), \qquad \forall \mathbf{h}\in \mathbb{R}^P,\forall \rho \in M_+(K_X).
\end{align}
where $\mathbf{h}_{+}=[\mathbf{h}]_{+}$ is the positive component of the vector $\mathbf{h}$. Using this inequality, we can find the minimum of $E(\mathbf{h};\rho)$ [as required in Eq.\eqref{eq:SIngle_neuron_condition}] via minimizing $E_r(\mathbf{h}_{+};\rho)$ over all the nonnegative vectors $\mathbf{h}_+ \geq 0$.

Minimizing $E_r(\mathbf{h}_{+};\rho)$ turns out to be equivalent to determining the copositivity of a matrix \cite{shaked2021copositive} and is generally a non-convex problem. Using the properties of the output kernel $K_Y$, we solve this non-convex problem and show that any solution $\rho_*$ must have the hidden representation kernel (SI \S 2.2)
\begin{equation}\label{eq:opt_kernel}
   K[\rho_*]=b_{*}(d_{Y}\mathbf{1}\mathbf{1}^{T}+K_{Y}). 
\end{equation}
where
\begin{equation}\label{eq:b_star}
b_{*}=\sqrt{\frac{\lambda_{2}}{\lambda_{1}}\frac{P+1}{P(P+2)}}-\frac{\lambda_{2}}{P}.
\end{equation}
Here $\lambda_{1,2}$ are assumed to be small to ensure $b_{*}>0$, i.e. $\lambda_{1}\lambda_{2}<\frac{P+2}{P(P+1)}$. Interestingly, the solutions of the mean-field equation $\rho_*$ [Eq.\eqref{eq:SIngle_neuron_condition}], or equivalently the global minimizers of the loss [Eq.\eqref{eq:loss_functional}], are not unique. But they all correspond to the identical representation kernel given by Eq.\eqref{eq:opt_kernel}. We note that such uniqueness of the representation kernel does not generally hold for arbitrary tasks and appears to be a special property of the tasks studied here (SI \S2.2)

\begin{figure}[tbp]
    \centering
    \includegraphics[width=\linewidth]{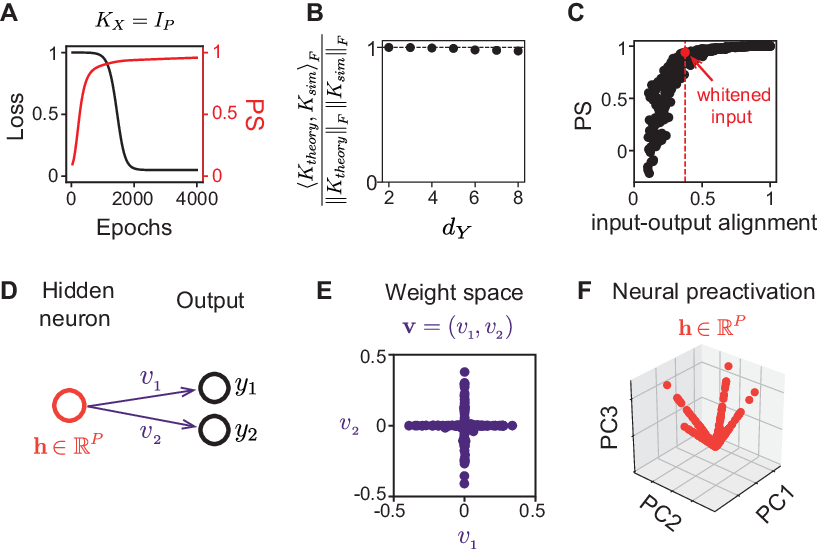}
    \caption{Task-optimized ReLU network exhibits abstract representation for whitened and target-aligned inputs. (A) The training loss and the parallelism score of the hidden representation are plotted against the number of training epochs for the whitened input. The training is through gradient descent algorithm. After training, the network performs the task perfectly with zero training loss and has an abstract hidden representation ($PS \rightarrow 1$). (B) The optimal hidden kernel predicted by theory ($K_{theory}$ given by Eq.~\eqref{eq:opt_kernel}) is aligned with the one found in numerical simulation ($K_{sim}$) for different output dimensions $d_Y$. (C) The parallelism score of the hidden representation after training vs. the input-output alignment. See SI \S 5 for the definition of input-output alignment. Each point in the plot represents a specific randomly sampled input geometry. The red dot indicates the point for the whitened input. For inputs that are more aligned to the output than the whitened one, the $PS$ is close to $1$. (D-E) Modularity of the single-neuron tuning in the hidden layer is captured by the preactivation vector $\mathbf{h}$ and weight vector $\mathbf{v}$ to the output layer for each hidden neuron ($d_Y = 2$). (E) Each hidden neuron only has nonzero output weights to a single output unit, suggesting that different neurons in the hidden layer "encode" different output labels. (F) The first three principal components of the neural preactivation space (the same space as Fig.~3B). The preactivation vectors of the hidden neurons concentrate along $P=4$ distinct directions, as predicted by Eq.~\eqref{eq:h_opt}.}
    \label{fig:2}
\end{figure}

The set of optimal preactivation vectors in the hidden layer $A(K_X,K_Y)$, which is equal to the support of $\rho_*$ , is found to consist of $2d_{Y}$ line rays in the nonnegative orthorant $\mathbb{R}_{\geq0}^{P}$ (SI \S 2.2)
\begin{align}
\mathbf{h}=\mathbf{h}_+=\alpha&(\mathbf{1}+\mathbf{v}_{i})\text{ or }\alpha(\mathbf{1}-\mathbf{v}_{i}), \nonumber \\ 
&\text{ for some }\alpha\geq0\text{ and }i \in \{1,2,..,d_{Y}\}.\label{eq:h_opt}
\end{align}
As half of the components in $\mathbf{v}_i$ are $+1$ and the other ones are $-1$, the above optimal preactivation vectors have exactly half of the components to be $2$ and the other ones zero. 

Finally, this optimal solution $\rho_{*}$ can be attained in a network with $M\geq 2d_{Y}$ hidden neurons where each neuron simply takes the preactivation $\mathbf{h}_{i}^{\pm}=\sqrt{b_{*}}(\mathbf{1}\pm\mathbf{v}_{i})$ (SI \S 2.2). 

Altogether, when $M\geq 2d_{Y}$ and the input is whitened, we find the optimal hidden representation is Eq.\eqref{eq:opt_kernel} and that the neurons in the hidden layer cluster into $2d_Y$ groups given by Eq.\eqref{eq:h_opt}.

\subsection{Whitened input + multi-element class ($n\protect\geq2$) \label{subsec:Orthogonal-input-+}}

When each class contains multiple data points ($n\geq2$ in Eq.~\eqref{eq:P_2dy}), whitened input has input kernel matrix $X_{data}^{T}X_{data}=I_{P}$. Namely, both the within-class and between-class correlations are zero. Here we consider a slightly more general form of input kernel that would allow within-class correlation,
\begin{equation}\label{eq:ortho_input_multi}
X_{data}^{T}X_{data}=\underset{n\cdot2^{d_{Y}}\times n\cdot2^{d_{Y}}}{\underbrace{\left(\begin{array}{cccc}
\underset{n\times n}{\underbrace{C_1}}\\
 & C_2\\
 &  & ..\\
 &  &  & C_{2^{d_Y}}
\end{array}\right)}} = \sum_{i=1}^{2^{d_Y}}C_i\otimes\mathbf{e}_i\mathbf{e}_i^T,
\end{equation}
where $\otimes$ is the Kronecker product between two matrices. $\mathbf{e}_i$, $i=1,2,..,2^{d_Y}$ is the standard basis in $\mathbb{R}^{2^{d_Y}}$. The two terms in the Kronecker product naturally represent the between-class and within-class correlations.

The within-class correlation matrices $C_i$ are assumed to satisfy the following conditions:
\begin{description}
\item [{(1)}] $C_i$ is positive definite for $i=1,2,3,..,2^{d_Y}$;
\item [{(2)}] All the $C_i$'s have the same largest eigenvalue $c>0$ with the same eigenvector $\mathbf{1}\in\mathbb{R}^{n}$, i.e. $C_i\mathbf{\mathbf{1}}=c\mathbf{1}$. 
\end{description}

Under this assumption, the augmented input kernel matrix is $K_X=\sum_{i=1}^{2^{d_Y}}C_i\otimes\mathbf{e}_i\mathbf{e}_i^T +\mathbf{1}\mathbf{1}^T \otimes \mathbf{1}\mathbf{1}^T$. And the output kernel becomes $K_Y = K_Y^0\otimes \mathbf{1}\mathbf{1}^{T}$. Here $K_Y^0$ denotes the $d_Y\times d_Y$ output kernel matrix in the single-element case. 

In SI \S 2.3, we solve the mean-field problem [Eq.~\eqref{eq:SIngle_neuron_condition}] and find the optimal representation kernel is similar as before
\begin{equation}\label{eq:opt_hidden_kernel_multi}
K[\rho_{*}]=b_{*}(d_{Y}\mathbf{1}\mathbf{1}^{T}+K_{Y}^{0})\otimes\mathbf{1}\mathbf{1}^{T},
\end{equation}
and $b_{*}$ is
\begin{equation}\label{eq:b_star_multi}
b_{*}=\sqrt{\frac{\lambda_{2}}{\lambda_{1}}\frac{P+c}{(P+2c)P}}-\frac{\lambda_{2}}{P}.
\end{equation}
Assume that $\lambda_{1,2}$ are small enough to have $b_{*}>0$. 

Similarly, the set of optimal neural preactivations $A(K_X,K_Y)$ is given by $2d_Y$ directions
\begin{widetext}
\begin{equation}\label{eq:Optimal_h_multielement}
A(K_{X},K_{Y})=\left\{ \mathbf{h}\in\mathbb{R}^{2^{d_{Y}}}\otimes\mathbb{R}^{n}\left|\mathbf{h}=\alpha(\mathbf{1}+\mathbf{v}^0_{i})\otimes\mathbf{1}\text{ or }\alpha(\mathbf{1}-\mathbf{v}^0_{i})\otimes\mathbf{1},\alpha\geq0\text{ and }i=1,2,..,d_{Y}\right.\right\},
\end{equation}
\end{widetext}
where $\mathbf{v}_i^0$'s are the eigenvectors of $K_Y^0$ with nonvanishing eigenvalues.

The optimal kernel [Eq.~\eqref{eq:opt_hidden_kernel_multi}] (and the associated optimal measure $\rho_{*}$) can be attained when the number of hidden neurons $M\geq 2d_{Y}$. Thus, despite having nonzero within-class correlations, the optimal kernel matrix is similar to the single-element case [Eq.~\eqref{eq:opt_kernel}].

This optimal representation kernel [Eq.~\eqref{eq:opt_hidden_kernel_multi}] has an intriguing geometric interpretation: the between-class correlation matrix $b_{*}(d_{Y}\mathbf{1}\mathbf{1}^{T}+K_{Y}^{0})$ is the same as for the single-element class case [Eq.~\eqref{eq:opt_kernel}] and the within-class correlation matrix is given by $\mathbf{1}\mathbf{1}^{T}$.  This form of within-class correlation means the neural representation of all data points from the same class ``collapse'' into a single point in the hidden layer, as also indicated in the set of optimal preactivations [Eq.~\eqref{eq:Optimal_h_multielement}]. Such a property closely resembles the neural collapse phenomena reported in previous studies \cite{papyan2020prevalence,zhu2021geometric,tirer2022extended,sukenik2023deep,sukenik2024neural,jacot2024wide}. We discuss the similarity and difference between our results and the work related to neural collapse in SI \S 5.1 in more detail. For ReLU nonlinearity, this "within-class collapse" property can be directly proved via Jensen's inequality (SI \S 2.4). The fact that the optimal preactivation patterns in the hidden layer clusters into only a few directions [Eq.~\eqref{eq:h_opt} and Eq.~\eqref{eq:opt_hidden_kernel_multi}] is also known as the quantization phenomenon \cite{maennel2018gradient}.


\subsection{Target-aligned input \label{subsec:Target-aligned-input}}

The above results for whitened input can be generalized to inputs more aligned to the output than the whitened case (Fig.~\ref{fig:2}C). We will state our main results for single-element $(n=1)$ and multi-element $(n\geq 2)$ classes here.

For the single-element class $(n=1)$, we consider the augmented input kernel matrix of the form
\begin{equation}\label{eq:Target_align_Kernel}
K_{X}=\frac{c_{0}}{P}\mathbf{1}\mathbf{1}^{T}+\frac{c_{Y}}{P}K_{Y}+\sum_{j=d_{Y}+1}^{P-1}c_{j}\mathbf{u}_{j}\mathbf{u}_{j}^{T},
\end{equation}
where $K_{Y}=\sum_{i=1}^{d_{Y}}\mathbf{\mathbf{v}}_{i}\mathbf{v}_{i}^{T}$ is the output kernel and $\mathbf{u}_{i}$'s are the orthonormal basis of $\text{span}\{\mathbf{1},\mathbf{v}_{1},\mathbf{v}_{2},...,\mathbf{v}_{d_{Y}}\}^{\perp}$ in $\mathbb{R}^{P}$ (Section \ref{subsec:Data-and-network}). 

The 2nd and 3rd terms in the above equation can be considered input components that are aligned and orthogonal to the output. For whitened inputs considered in Section \ref{subsec:Whitened-input-+}, $c_{0}=P+1,c_{Y}=c_{j}=1$, $j=d_{Y}+1..,P-1$. Here we consider those inputs whose output-aligned component is greater than or equal to the orthogonal component, meaning that 
\begin{equation}\label{eq:TargetAlignCondition}
c_{0}>c_{Y}\geq c_{j}>0, \qquad j=d_{Y}+1,..,P-1.
\end{equation}

Unlike the fully orthogonalized input before, the input here has positive (between-class) correlations if $c_{0}$ is large. Interestingly, we find (SI \S 2.6) that this introduced correlation does not change the form of the optimal kernel $K[\rho_*]=b_*(d_Y\mathbf{1}\mathbf{1}^T + K_Y)$, and only modifies the coefficient $b_*$
\begin{equation}\label{eq:b_star_targetAlign}
b_{*}=\sqrt{\frac{\lambda_2c_0c_YP}{\lambda_1[P(c_0-c_Y)+2c_Y]}} - \sqrt{\frac{\lambda_2}{P}}
\end{equation}

The same result holds for multi-element classes ($n\geq2$, $P=n\cdot 2^{d_{Y}}$). Now the augmented input kernel has the tensor product form 
\begin{align}\label{eq:Target_aligned_multi}
K_{X} = K_{X}^{0}\otimes C+\mathbf{1}\mathbf{1}^{T}\otimes\mathbf{1}\mathbf{1}^{T}
\end{align}
where $K_X^0$ is the input kernel in the single-element class scenario [Eq.~\eqref{eq:Target_align_Kernel}] and satisfies Eq.~\eqref{eq:TargetAlignCondition}. For simplicity, we also assume that all the classes have the same within-class correlation matrix $C$, which satisfies the same properties as before (Section \ref{subsec:Whitened-input-+}). This input kernel recovers the one for orthogonalized classes (Section \ref{subsec:Orthogonal-input-+}) if $c_{0}=1,c_{Y}=c_{j}=1$, and recovers single-element class [Eq.~\eqref{eq:Target_align_Kernel}] if every class has a single element $(n=1)$. An example dataset with the above input kernel is when the within-class correlation is $c_{in}$ and between-class correlation is $c_{out}$, with $c_{out} < c_{in}$.

The optimal kernel in this case has the same form as for orthogonalized inputs [Eq.~\eqref{eq:opt_hidden_kernel_multi}], $K[\rho_{*}]=b_{*}(d_{Y}\mathbf{1}\mathbf{1}^{T}+K_{Y}^{0})\otimes\mathbf{1}\mathbf{1}^{T}$, but with an coefficient (SI \S 2.6)
\begin{equation*}
b_{*}= \sqrt{\frac{\lambda_{2}}{\lambda_{1}}\frac{cc_Y(P+cc_0)}{(P+cc_0+cc_Y)P}}-\frac{\lambda_{2}}{P}.
\end{equation*}

This optimal kernel (and the associated optimal measures $\rho_{*}$) can be attained when the number of hidden neurons $M\geq 2d_{Y}$. The set of optimal preactivations $A(K_X,K_Y)$ is the same as before [Eq.~\eqref{eq:h_opt} or Eq.~\eqref{eq:Optimal_h_multielement}].

Altogether, this shows that for target-aligned inputs [Eq.\eqref{eq:TargetAlignCondition}], the optimal representation kernel [Eq.~\eqref{eq:Target_align_Kernel}] does not depend on the magnitude of orthogonal components $c_{j}$ for all $j=d_Y+1,..,P-1$. However, this property does not hold if the target-aligned condition [Eq.\eqref{eq:TargetAlignCondition}] is not satisfied, i.e. when the orthogonal component is large, $c_j>c_Y$ (Fig.\ref{fig:3}C). The underlying intuition is that for nonlinear networks, the output-aligned component and the output-orthogonal component in the input "compete" with each other to form the hidden representation. In contrast, a linear network only uses the output-aligned component to form the hidden representation.


\subsection{$PS$, single-neuron tuning and weight matrices for the optimal solution\label{subsec:,-single-neuron-tuning}}

We have shown that for orthogonalized [Eq.~\eqref{eq:ortho_input_multi}] or target-aligned [Eq.~\eqref{eq:Target_aligned_multi}] inputs, when $M\geq2d_{Y}$, all global minima of the loss have the representation kernel $K[\rho_{*}]=b_{*}\left(d_{Y}\mathbf{1}\mathbf{1}^{T}\otimes\mathbf{1}\mathbf{1}^{T}+K_{Y}\right)$. In this section, we investigate other properties of these optimal solutions.

From the optimal representation kernel, we can compute its $PS$, denoted $PS(K[\rho_{*}])$. By translation-and scale-invariance of $PS$ (Section \ref{subsec:Parallelism-score-measures}), $PS(K[\rho_{*}])=PS(K_{Y})$. 
By the definition of $K_{Y}$, the prototype representation [Eq.~\eqref{eq:prototype_rep}] of each binary class corresponds to each vertex of a $d_{Y}$-dimensional hypercube (Fig.~\ref{fig:1b}). 
So the representation change for the $k$th binary latent label is aligned with the $k$th axis of the hypercube (Fig.\ref{fig:2}), independent of other latent labels: $PS_{k}(\alpha_{1},\alpha_{2})\equiv\frac{\Delta\mathbf{r}(k;\alpha_{1})\cdot\Delta\mathbf{r}(k;\alpha_{2})}{\left\Vert \Delta\mathbf{r}(k;\alpha_{1})\right\Vert _{2}\left\Vert \Delta\mathbf{r}(k;\alpha_{2})\right\Vert _{2}}=1$. Therefore, the overall parallelism score $PS(K_{Y})=1=PS(K[\rho_{*}])$, and the optimal representation kernel [Eq.~\eqref{eq:ortho_input_multi} and \eqref{eq:Target_aligned_multi}] corresponds to an abstract representation.

\begin{figure*}[tbp]
    \centering
    \includegraphics[width=0.6\linewidth]{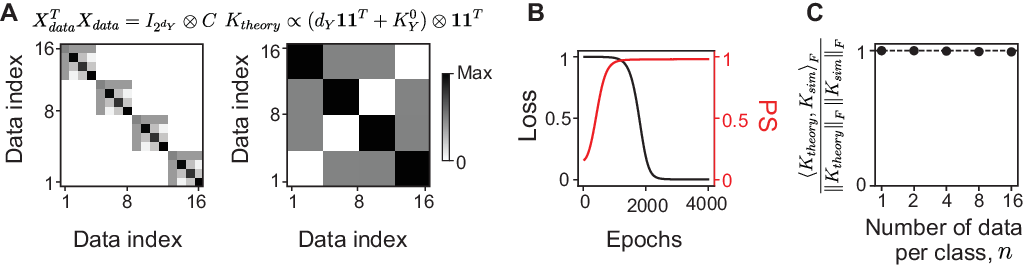}
    \caption{Task-optimized ReLU network for multi-element classes ($n\geq 2$). (A) An example of input kernel with nonzero within-class correlation but zero between-class correlation (left). For illustration, this example assumes the same within-class correlation matrix $C$ for all classes but our theory also works for class-specific correlation [Eq.\eqref{eq:ortho_input_multi}]. The theory predicts that the optimal hidden kernel $K_{theory}$ is proportional to the output kernel up to a positive shift (right). In particular, the hidden kernel has a block structure because all the data within each binary class has the same hidden representation [Eq.\eqref{eq:Optimal_h_multielement}]. (B) Training loss and $PS$ are plotted against the number of training epochs. (C) The predicted hidden kernel ($K_{theory}$) always aligns with simulation ($K_{sim}$) when the number of data per class $n$ changes.}
    \label{fig:3}
\end{figure*}

Next, we look at the responses of individual neurons in the optimal solution $\rho_{*}$. From Eq.~\eqref{eq:h_opt} and Eq.~\eqref{eq:Optimal_h_multielement}, the optimal preactivation of each neuron has the form $\mathbf{h}=\alpha(1\pm\mathbf{v}^0_{i})\otimes\mathbf{1}$ for some $i\in\{1,2,3..,d_{Y}\}$ and $\alpha>0$ (we focus on neurons having nonzero preactivation patterns). The $P$ components of $\mathbf{h}\in\mathbb{R}^{P}\simeq\mathbb{R}^{2^{d_{Y}}}\otimes\mathbb{R}^{n}$ are this neuron's preactivations for the $P$ training data points [Eq.\eqref{eq:preact_matrix}]. Based on the definition of $\mathbf{v}^0_{i}$ (Section \ref{subsec:Data-and-network} and SI \S 1.1), the neuron with $\mathbf{h}=\alpha(1+\mathbf{v}^0_{i})\otimes\mathbf{1}$ (or $\mathbf{h}=\alpha(1-\mathbf{v}^0_{i})\otimes\mathbf{1}$) has nonzero preactivations exactly for those training data points whose $i$th output label are positive (or negative). So all the neurons in the hidden layer are divided into $2d_{Y}$ groups: each group responds only to a single output label (Fig.\ref{fig:3}D-F). Note that a random network would not have a modular response property like this, and such modularity is a consequence of task training.

Since the preactivation of every neuron is a linear combination of $\mathbf{1\otimes\mathbf{1}}$ and $\mathbf{v}_i\otimes\mathbf{1}$ for some $i$, the optimal preactivation matrix $H_{*}$ [Eq.~\eqref{eq:preact_matrix}] has the following form
\begin{equation*}
H_{*}=\underset{P\times (d_Y+1) \text{ matrix}}{\underbrace{\left(\mathbf{1\otimes\mathbf{1}},\mathbf{v}_{1}\otimes\mathbf{1},\mathbf{v}_{2}\otimes\mathbf{1},..,\mathbf{v}_{d_{Y}}\otimes\mathbf{1}\right)}}\mathbf{A},
\end{equation*}
where $\mathbf{A}\in\mathbb{R}^{(d_Y+1)\times M}$ is the matrix encoding the coefficients $\alpha$ for all neurons. Importantly, $H_{*}\in\mathbb{R}^{P\times M}$ has rank $d_{Y}+1$. The corresponding optimal weight matrices for Eq.~\eqref{eq:TwoLayerNN_loss} can be solved as $W_{*}=H_{*}X^\dagger$ (since $X$ is of full rank) and $W_{2,*}=YH_{*}^{T}[\lambda_{2}+H_{*}H_{*}^{T}]^{-1}$ (SI \S 1.2). So these matrices also have rank $d_{Y}+1$, which could be much smaller than the hidden layer width $M$ and input dimensions $d_{X}$. The fact that the optimal weight matrices are low-rank indicates that the neural network learns the task-relevant low-dimensional structures in the input, rather than constructs a large set of fixed features from the inputs as in kernel machine \cite{pmlr-v125-woodworth20a,farrell2023lazy}.

\section{Abstract representation emerges in the hidden layer, independent of single-neuron nonlinearity\label{sec:Abstract-representation-emerges}}

We show that the results from the previous section can be extended to other activation functions $\phi$ beyond ReLU. In particular, the emergence of abstract representation is robust to single-neuron nonlinearity and is mainly determined by the task structure. 

We consider two broad classes of nonlinear activation functions. The 1st class is the threshold nonlinear functions of the form 
\begin{equation}\label{eq:1st_type_nonlinear}
\phi(z)=\begin{cases}
\phi_{+}(z) & z\geq0\\
0 & z<0
\end{cases}
\end{equation}
Here we require the nonzero function $\phi_{+}:\mathbb{R}_{\geq 0}\rightarrow\mathbb{R}_{\geq 0}$ to satisfy the following properties: 
\begin{description}
\item [{(1)}] $\phi_{+}$ is continuous, non-decreasing and $\phi_{+}(0)=0$;
\item [{(2)}] For any $z>0$, the slope function $B(z)\equiv\frac{\phi_{+}(z)}{z}\geq0$ is non-increasing, and $B_{0}\equiv\underset{z\rightarrow0^{+}}{\lim}B(z)\in(0,+\infty)$. 
\end{description}
The property \textbf{(2)} can be viewed as a saturation effect in the neural responses. Examples in this function class include ReLU, hard Sigmoid, and functions that are concave for positive inputs (Fig.~\ref{fig:4}A). 

The 2nd class of nonlinear activations is those odd functions,
\begin{equation}
\phi(z)=\begin{cases}
\phi_{+}(z) & z\geq0\\
-\phi_{+}(-z) & z<0
\end{cases},\label{eq:2nd_type_nonlinear}
\end{equation}
where $\phi_{+}$ satisfies the same properties \textbf{(1)} and \textbf{(2)} as above. This class of functions includes linear functions, tanh, and any odd function that is concave for positive inputs (Fig.~\ref{fig:5}A).

In SI \S 3, we find that for orthogonalized and target-aligned inputs, the optimal representation kernel converges to the form $K[\rho_{*}^{M}]\rightarrow a_{*}\mathbf{1}\mathbf{1}^{T}\otimes\mathbf{1}\mathbf{1}^{T}+b_{*}K_{Y}$ as $M\rightarrow+\infty$. As noted before, this optimal kernel corresponds to an abstract neural representation. 

Our central strategy is to (i) prove these results for a perturbed version of the nonlinear function, and then (ii) extend the result via a continuity argument. We outline these key arguments below for the single‐element class case ($n=1$) with whitened input. Detailed derivations, as well as the scenarios with multi‐element classes ($n \ge 2$) and target‐aligned inputs, are provided in SI \S 3.


\subsection{Threshold nonlinear activation\label{subsec:Threshold_nonlinear}}
For any $\delta>0$ and a 1st class nonlinearity $\phi$, we introduce the perturbed activation function
\begin{equation*}
\phi^{\delta}(z)\equiv\begin{cases}
B_{0}\delta - \phi_+(\delta)+\phi_{+}(z) & z\geq\delta\\
B_{0}z & 0\leq z\leq\delta\\
0 & z<0
\end{cases},
\end{equation*}
where $B_{0}=\underset{z\rightarrow0^{+}}{\lim}\frac{\phi_{+}(z)}{z}>0$ is the slope of $\phi$ at the origin. Here $\phi$ is simply replaced by its linear tangent on $[0,\delta]$. Moreover, it can be checked that if $\phi$ is a 1st class nonlinearity, so is the perturbed one $\phi^{\delta}$.

For this perturbed activation $\phi^{\delta}$ and whitened input $K_{X}=I_{P}+\mathbf{1}\mathbf{1}^{T}$ (single-element class $n=1$), we find (SI \S 3.1-3.2) that a similar result as Eq.~\eqref{eq:whiten_1stterm} holds
\begin{equation}\label{eq:1stterm_1stClass}
     E(\mathbf{h};\rho) \geq E_{r}(\phi^{\delta}(\mathbf{h});\rho), \quad \forall \mathbf{h}\in \mathbb{R}^P,\forall \rho \in M_+(K_X),
\end{equation}
where $E_{r}(\cdot;\rho)$ is given by Eq.~\eqref{eq:whiten_1stterm} except that the regularization strength $\lambda_1$ is rescaled $\lambda_{1}\rightarrow\lambda_{1}B_{0}^{-2}$. This result can be proved via the method of majorization \cite{bhatia2013matrix,marshall1979inequalities} (SI \S 3.1-3.2).

Note that the argument $\phi^{\delta}(\mathbf{h})$ in $E_r(\phi^{\delta}(\mathbf{h});\rho)$ takes values from a hypercube $\phi^{\delta}(\mathbf{h})\in[0,\phi_{+}(+\infty)]^{P}$ within the nonnegative orthant $\mathbb{R}^P_{\geq0}$. Minimizing $E_{r}(\phi^{\delta}(\mathbf{h});\rho)$ is a nonconvex copositive programming problem similar to the ReLU case (SI \S 3.2), and we find that the optimal representation kernel is still of the form $K^{\delta}[\rho_{*}]=b_{*}(d_{Y}\mathbf{1}\mathbf{1}^{T}+K_{Y})$, with $b_{*}$
\begin{equation*}
 b_{*}=\sqrt{\frac{\lambda_{2}B_{0}^{2}}{\lambda_{1}}\frac{P+1}{P(P+2)}}-\frac{\lambda_{2}}{P},
\end{equation*}
where $\lambda_{1,2}$ are assumed to be small enough such that $b_{*}>0$. So the effect of the nonlinearity $\phi^{\delta}$ is to simply scale $\lambda_{1}$ by $B_0^{-2}$, the inverse square of the slope of $\phi_+$ at the origin.

The set of optimal preactivations in this case is the same as the ReLU case [Eq.~\eqref{eq:h_opt}] but with an additional constraint $\mathbf{h}=B_0^{-1}\phi^{\delta}(\mathbf{h})$ (SI \S 3.2). Furthermore, this optimal solution is attained when the hidden layer has a width $M> 2d_Y\left \lceil \sqrt{b_*}\delta^{-1} \right \rceil$ (SI \S 3.2).

Note that the optimal kernel (and $b_*$) is independent of $\delta >0$, so we can carry out the limiting argument $\delta\downarrow0^{+}$ and find that the same kernel $K^{\delta}[\rho_{*}]=b_{*}(d_{Y}\mathbf{1}\mathbf{1}^{T}+K_{Y})$ is still optimal even if $\delta = 0$ (SI \S 3.4).

\begin{figure}[tbp]
    \centering
    \includegraphics[width=\linewidth]{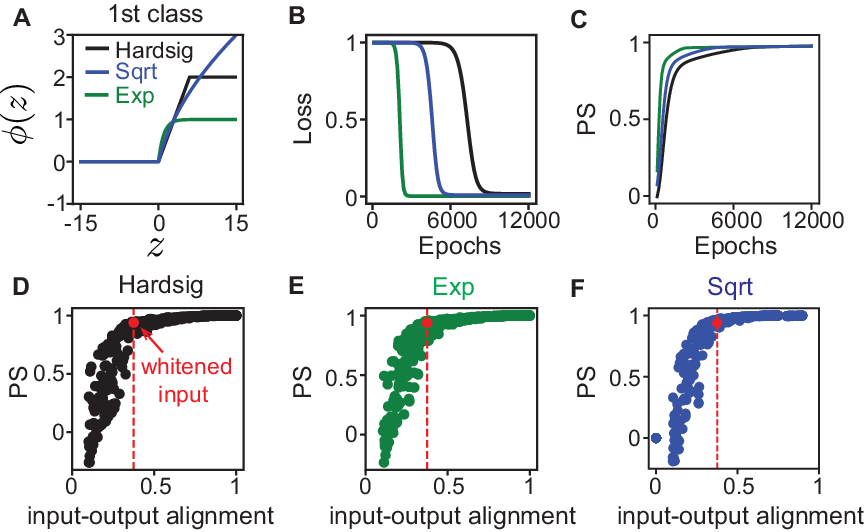}
    \caption{Task-optimized network with threshold nonlinearity. (A) Three examples of nonlinear activation functions from the 1st class. See detailed expressions for the three functions in SI \S 5. (B,C) Training loss and $PS$ of the hidden representation are plotted against the number of training epochs, for the three activation functions. (D-F) $PS$ of the optimal hidden representation when varying the input-output alignments. This plot appears to be insensitive to the specific shape of nonlinearity used in the network.}
    \label{fig:4}
\end{figure}

While these results show that for whitened and target-aligned inputs, the optimal representation for any 1st class nonlinear activation function $\phi$ is always abstract, interestingly, we find in simulations that the $PS$ of the optimal representation for other input geometries seems to be also robust to the choice of nonlinearity for a wide network (Fig.\ref{fig:4}D-F).

\subsection{Odd-symmetric nonlinear activation}

Similarly, for the 2nd class of nonlinearity, the perturbed activation function is
\begin{equation*}
\phi^{\delta}(z)\equiv\begin{cases}
B_{0}\delta- \phi_+(\delta)+\phi_{+}(z) & z\geq\delta\\
B_{0}z & -\delta\leq z\leq\delta\\
-B_{0}\delta + \phi_+(\delta) -\phi_{+}(-z) & z\leq-\delta
\end{cases}
\end{equation*}
where $B_{0}$ is the slope at $z\rightarrow0^{+}$. If $\phi$ is a 2nd class nonlinearity, so is the perturbed one $\phi^{\delta}$.

For single-element class ($n=1$) and whitened inputs, a similar bound as Eq.~\eqref{eq:1stterm_1stClass} holds (SI \S 3.3)
\begin{equation*}
    E(\mathbf{h};\rho) \geq E_{r}(\phi^{\delta}(\mathbf{h});\rho), \quad \forall \mathbf{h}\in \mathbb{R}^P,\forall \rho \in M_+(K_X),
\end{equation*}
where $E_r(\cdot;\rho)$ is given by Eq.~\eqref{eq:whiten_1stterm} except that the regularization strength $\lambda_1$ is rescaled $\lambda_{1}\rightarrow\lambda_{1}B_{0}^{-2}$.

Note that there is one crucial difference for the 2nd class nonlinearity compared to the 1st class nonlinearity: the argument in $E_{r}(\phi^{\delta}(\mathbf{h});\rho)$, $\phi^{\delta}(\mathbf{h})\in[-\phi_{+}(+\infty),\phi_{+}(+\infty)]^{P}$ takes values within the hypercube region that is symmetric to the origin. Particularly, it can take negative values. Due to this difference, minimizing $E_r(\phi^{\delta}(\mathbf{h});\rho)$ becomes a convex quadratic programming problem rather than a nonconvex copositive programming problem.

Specifically, the unique optimal kernel in this case is found to be
\begin{equation}\label{eq:opt_ker_2ndClass}
    K^{\delta}[\rho_{*}]=b_{*}K_{Y}.
\end{equation}
where 
\begin{equation*}
b_{*}=\sqrt{\frac{\lambda_{2}B_{0}^{2}}{\lambda_{1}P}}-\frac{\lambda_{2}}{P}.
\end{equation*}
$\lambda_{1,2}$ are small enough to ensure $b_{*}\geq0$. We note that there is no constant shift (the $\mathbf{1}\mathbf{1}^T$) in the above kernel.

Moreover, the set of optimal preactivations (SI \S 3.3) forms a subspace, having the form
\begin{align}\label{eq:opt_preactivation_2nd_class}
\mathbf{h}=\sum_{i=1}^{d_{Y}}\alpha_{i}\mathbf{v}_{i}\text{, }\alpha_{i}\in\mathbb{R}\text{ and }i\in \{1,2,..,d_{Y}\},
\end{align}
with the additional constraint $\mathbf{h}=B_{0}^{-1}\phi^{\delta}(\mathbf{h})$. Furthermore, this optimal solution can be attained when the network width $M> d_Y\left \lceil b_*\delta^{-1} \right \rceil$.

Despite the differences, the optimal kernel in Eq.~\eqref{eq:opt_ker_2ndClass} still represents an abstract representation. Finally, repeating the limiting argument as before shows that this representation kernel remains optimal when taking $\delta\downarrow 0^{+}$ (SI \S 3.4).

\subsection{Differences in single-neuron tuning}

\begin{figure}[tbp]
    \centering
    \includegraphics[width=\linewidth]{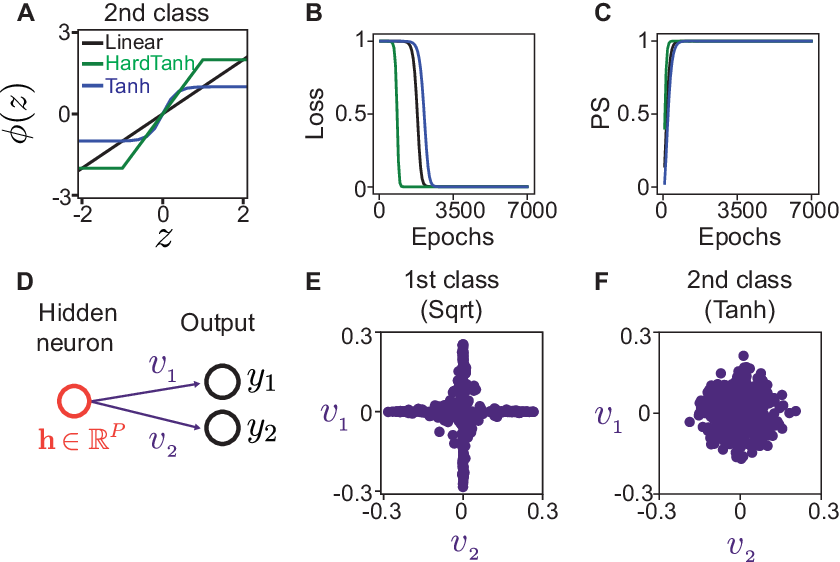}
    \caption{Task-optimized network with odd-symmetric nonlinearity. (A) Three examples of nonlinear activation functions from the 2nd class. See detailed expressions for the three functions in SI \S 5. (B,C) Training loss and $PS$ of the hidden representation as a function of the number of training epochs, for the three activation functions in (A). (D-F) Although the optimal hidden representations are always abstract for the two classes of nonlinearity, the hidden neurons have different tuning properties. For 1st class nonlinearity, each hidden neuron is specifically connected to a single output unit (E), while this connectivity appears unstructured for hidden neurons when using the 2nd class nonlinearity (F).}
    \label{fig:5}
\end{figure}

For both classes of nonlinearity [Eq.~\eqref{eq:1st_type_nonlinear}-\eqref{eq:2nd_type_nonlinear}], we have shown that the optimal hidden representation in the network is abstract and has $PS$ equal to $1$. While the abstractness of the representation is robust to different choices of nonlinearity, the single-neuron tuning properties are different for two classes of nonlinearity (Fig.~\ref{fig:5}D-F). 

When the activation function is ReLU (which belongs to the 1st class of nonlinearity [Eq.~\eqref{eq:1st_type_nonlinear}]), we have shown that the optimal neural representation consists only of neurons that are either positively or negatively tuned to a single output label (Section \ref{subsec:,-single-neuron-tuning}), yielding  $2d_{Y}$ groups of neurons. For general 1st class nonlinearity $\phi$, since it shares the same set of optimal preactivations [Eq.~\eqref{eq:h_opt}] as ReLU, this modular tuning remains \footnote{More precisely, we show that in SI \S 3.1-3.2, the perturbed nonlinearity shares the same set of optimal preactivations as ReLU. By continuity, this form of modularity persists when $\delta\downarrow0^{+}$.} (Fig.~\ref{fig:5}E). 

On the other hand, for odd activation functions [Eq.~\eqref{eq:2nd_type_nonlinear}] (including the linear activation function), the optimal preactivations [Eq.~\eqref{eq:opt_preactivation_2nd_class}] can freely rotate within a subspace (when $\mathbf{h}$ is small). So in this case, neurons in the hidden layer generally exhibit mixed selectivity \cite{rigotti2013importance,fusi2016neurons} (Fig.~\ref{fig:5}F), where many components of $\mathbf{h}$ are nonzero. 

Therefore, even if both classes of nonlinearity generate the same optimal population geometry (characterized by the same representation kernel $K[\rho_{*}]$ up to a global shift), they lead to different single-neuron tuning properties. This suggests that the optimal tuning curves on the single-neuron level are not only impacted by the task structure, but also by the biophysical properties of individual neurons (which determine the response nonlinearity).

\section{Extensions\label{sec:Extensions-and-other}}

In the previous sections, we use our analytical framework to study optimal neural representations for two-layer networks trained on the data model in Section \ref{subsec:Data-and-network}. However, the framework is indeed applicable to additional data models and deep neural networks. For illustrative purposes, we present these extensions here, with a focus on the single-element class ($n=1$) with ReLU nonlinear activation. These results can be extended to any nonlinearity as before (Section \ref{sec:Abstract-representation-emerges}).

We first discuss some general properties of the optimal representation kernel derived using our analytical framework (Section \ref{sec:The-analytical-framework}). Given any input and output matrices $X,Y$, denote the set of optimal representation kernels as $S(X,Y)\subseteq \mathbb{R}^{P \times P}$. According to Eq.~\eqref{eq:loss_h}, the optimal preactivation matrix $H$ and thus the optimal representation kernel only depend on the input and output kernels, $S(X,Y)\equiv S(K_{X},K_{Y})$. Moreover, this set is invariant under global scaling factors of the input and output kernels (for $\phi=$ ReLU and small $\lambda_{1,2}$), $S(\alpha_{X}K_{X},\alpha_{Y}K_{Y})=S(K_{X},K_{Y})$ if $\alpha_X,\alpha_Y>0$. 

\textbf{(1)} Denote the set of optimal measures of Eq.~\eqref{eq:loss_functional} as $\mathcal{P}(K_{X},K_{Y})\subseteq M_{+}(X)$. Since the problem is a convex optimization problem, the optimal solution set, $\mathcal{P}(K_{X},K_{Y})$ is a convex set and in particular, is simply connected. Because the kernel $K[\cdot]$ is a linear map on $M_{+}(X)$, the set of optimal representation kernel matrices $S(K_{X},K_{Y})=K[\mathcal{P}(K_{X},K_{Y})]$ is also convex and simply connected. For ReLU nonlinearity, $S(K_{X},K_{Y})$ is a subset of the set of completely positive matrices and by Caratheodory's theorem, can be attained when the network width $M \geq 1+ P(P+1)/2$ \cite{shaked2021copositive}. This property is usually called mode connectivity of the loss landscape \cite{freeman2016topology,bach2017breaking,nguyen2019connected,simsek2021geometry}, and our analytical framework provides an alternative way to uncover this property.

\textbf{(2)} For input kernel $K_{X}$, denote $K[\rho_{*}]\in S(K_{X},K_{Y})$ as the optimal kernel with the set of optimal preactivations $ A(K_{X},K_{Y})$. Now if a rank one perturbation is added to the input kernel, $K_{X}^{new}=K_X-a\mathbf{v}\mathbf{v}^{T},\, a\in \mathbb{R},\mathbf{v}\in \text{Range}K_X$, and moreover, its magnitude and direction satisfy
\begin{equation*}
    a \mathbf{v}^TK_X^{\dagger}\mathbf{v} <1 \quad \text{and} \quad K^{\dagger}_X\mathbf{v} \in A(K_X,K_Y)^{\perp},
\end{equation*}
then the same kernel $K[\rho_{*}]$ is also optimal for this perturbed input (SI \S 4.1). This property shows that the optimal hidden representation is robust to input perturbation along certain directions.

\textbf{(3)} If the pseudo-inverse of the input kernel $K_X^{\dagger}$ has negative off-diagonals, we can show that the optimal set of neural preactivations lies entirely within the nonnegative orthorant, $A(K_X,K_Y)\subseteq \mathbb{R}^P_{\geq0}$. And this property holds independently of the output kernel $K_Y$. In this specific scenario, the problem of training two-layer network [Eq.~\eqref{eq:TwoLayerNN_loss}] is equivalent to a dictionary learning problem with positive representations. Input kernels $K_{X}$ satisfying such properties are known as inverse M-matrix \cite{fiedler1962matrices,johnson1982inverse,johnson2011inverse,dellacherie2014inverse} (SI \S 4.2).

\textbf{(4)} For any optimal measure $\rho_* \in \mathcal{P}(K_X,K_Y)$, the output of the network for a new data point $\mathbf{x}=(\mathbf{x}_{data},1)^T\in \mathbb{R}^{d_X+1}$ is (SI \S 4.3),
\begin{equation*}
    \quad \mathbf{y}_*(\mathbf{x}) =Y\frac{1}{\lambda_{2}+K[\rho_{*}]}\int\phi(\mathbf{h})\phi(\mathbf{h}^{T}K_{X}^{\dagger}X^{T}\mathbf{x})d\rho_{*}(\mathbf{h}),
\end{equation*}
where $X$ is the augmented input matrix [Eq.~\eqref{eq:aug_matrix}] and $Y$ is the output matrix [Eq.~\eqref{eq:input_output_mat}]. This result thus allows us to investigate the generalization property of the optimal network solution. As an illustration, we apply it to a simple compositional‐generalization task (see SI \S 4.4).

\subsection{Anisotropic input-output geometry}

In this section, we consider a scenario where the input and output kernels are
\begin{align*}
K_{X}&=\frac{c_{0}}{P}\mathbf{1}\mathbf{1}^{T}+\sum_{i=1}^{d_{Y}}\frac{c_{i}}{P}\mathbf{v}_{i}\mathbf{v}_{i}^{T}+\sum_{j=d_{Y}+1}^{P-1}c_{j}\mathbf{u}_{j}\mathbf{u}_{j}^{T},\nonumber \\
K_{Y}&=\sum_{i=1}^{d_{Y}}d_{i}\mathbf{v}_{i}\mathbf{v}_{i}^{T}.
\end{align*}
According to the definition of $\mathbf{v}_i$ (Section \ref{subsec:Data-and-network}), the above output kernel would correspond to anisotropic output labels where the $i$th output direction is scaled by a factor $d_{i}>0$ for different $i$'s, yielding a hierarchical structure (Fig.~\ref{fig:6}A). We assume that the inputs are also anisotropic ($c_{i}>0$'s are different) and are target-aligned,
\begin{align*}
c_{0}&>\underset{i=1,..d_{Y}}{\text{max}}c_{i}, \quad\text{and}\quad \underset{i=1,..d_{Y}}{\text{min}}c_{i}\geq\underset{j=d_{Y}+1,..P-1}{\text{max}}c_{j}.
\end{align*}
This kernel recovers scenarios in previous sections if $d_i=1,c_i=c_Y$ for all $i$'s.

We solve the mean-field problem [Eq.~\eqref{eq:SIngle_neuron_condition}] in this case and find that the unique optimal hidden representation kernel is (SI \S 4.5)
\begin{equation*}
K[\rho_{*}]=\sum_{i=1}^{d_{Y}}b_{i}^{*}\left(\mathbf{1}\mathbf{1}^{T}+\mathbf{v}_{i}\mathbf{v}_{i}^{T}\right)
\end{equation*}
where the coefficient $b^*$ is
\begin{equation*}
     b_{i}^{*}= \sqrt{\frac{\lambda_2d_ic_0c_iP}{\lambda_1(c_0+c_i)}} - \sqrt{\frac{\lambda_2}{P}}.
\end{equation*}
We see that up to a global translation, the effect of anisotropic input and output simply scales the $i$th direction of the hidden representation by a factor $b_i^*$ that depends on $d_i$. The hidden representation still has a hypercube (or hyper-rectangular) geometry (Fig.~\ref{fig:6}C) and is abstract. However, the anisotropy in the training data induces a more pronounced stage-like transition in the learning dynamics (Fig.~\ref{fig:6}B).

\subsection{Deep neural network \label{subsec:Deep-neural-network}}

\begin{figure}[tbp]
    \centering
    \includegraphics[width=\linewidth]{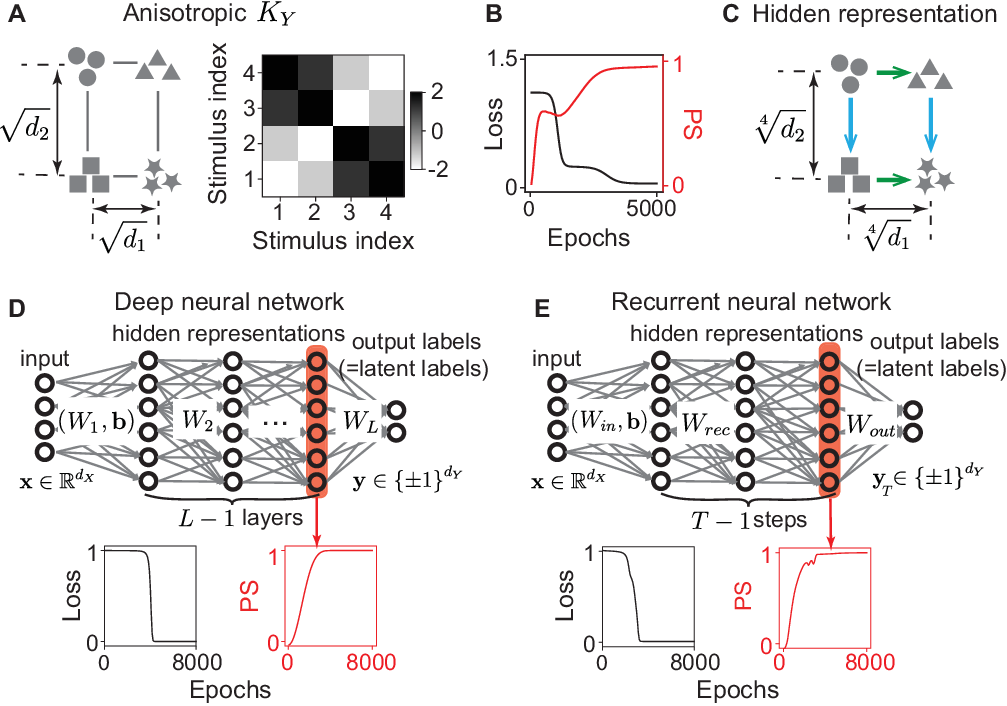}
    \caption{Extensions of the analytical framework to anisotropic input-output geometry, deep feedforward network, and recurrent network models. (A) For anisotropic output geometry, different output dimensions are scaled by a different factor $\sqrt{d_i}$. The corresponding output kernel matrix is shown to have a hierarchical block structure. (B) Training loss and $PS$ of hidden representation are plotted against the number of training epochs. The training dynamics has more prominent stage-like transitions than the case with isotropic outputs in Figure \ref{fig:2}A. (C) The optimal hidden representation is aligned with the output geometry in (A), except that the axis representing each latent label is rescaled. It has $PS$ equal to $1$. (D) A deep neural network trained on related tasks develops an abstract representation in its last layer. (E) A recurrent neural network trained on the related tasks develops an abstract representation at the last timestep.}
    \label{fig:6}
\end{figure}

The analytical framework presented in Section \ref{sec:The-analytical-framework} generalizes naturally to deep networks (Fig.~\ref{fig:6}D). We consider a deep feedforward network
\begin{equation*}
f_{\{W^{l}\}}(\mathbf{x})=W^{L}\phi(W^{L-1}\phi(..\phi(W^1\mathbf{x}))),
\end{equation*}
For convenience, we have incorporated the bias parameter of the 1st layer into the input $\mathbf{x}$, and the rest of the layers are bias-free. The loss function is
\begin{equation*}
E(\{W^{l}\})\equiv\left\Vert Y-f_{\{W^{l}\}}(X)\right\Vert _{F}^{2}+\sum_{l=1}^{L}\lambda_{l}\left\Vert W^{l}\right\Vert _{F}^{2}.
\end{equation*}
We are interested in when the last layer of the network exhibits abstract representations (Fig.~\ref{fig:6}D) for small regularization parameters $\lambda_{l}\ll1,l=1,2,..,L$.

Following similar procedures as in Section \ref{subsec:The-effective-mean-field}, we can introduce the empirical measure for preactivations for each layer $\rho^{l}=\sum_{k=1}^{M}\delta_{\mathbf{h}_k^{l}}$ and derive the KKT conditions for the optimal solutions (SI \S 4.6). 

For the data model in Section \ref{subsec:Data-and-network} with whitened input $K_{X}=I_{P}+\mathbf{1}\mathbf{1}^{T}$, we solve these KKT conditions to get an optimal representation kernel of the form
\begin{equation*}
K[\rho_{*}^{l}]=b_{*}^{l}(d_Y\mathbf{1}\mathbf{1}^{T}+K_{Y}),\qquad l=1,2,..,L.
\end{equation*}
In the limit $\lambda_{l}\equiv \lambda \downarrow 0^+$, $l=1,2,..,L$, these coefficients are
\begin{align*}
b^{1}_* & =\gamma_*\frac{(d_Y+1) (P+1)}{d_YP(P+2)},\nonumber \\
b^{l}_* & =(\gamma_*)^{l-1}b^1_*,\quad l=2,...,L-1.
\end{align*}
where $\gamma_*= \sqrt[L]{\frac{d_Y^2P(P+2)}{(d_Y+1)^2(P+1)}} + O(\lambda_l)$. The above result gives an optimal network that exhibits abstract representation in the last layer (and all the other layers, Fig.~\ref{fig:6}D) and can be attained when the width of every layer $M\geq 2d_Y$ (SI \S 4.6). In general, the effective energy function for deep networks ($L\geq 3$) is not convex over the space of measures [unlike Eq.\eqref{eq:loss_convexity}]. But the above solution is always a strict local minimum of the loss for any number of hidden layers $L \geq 2$ (SI \S 4.6). 

Finally, the analytical framework is extended to analyze the optimal representation in recurrent neural networks (SI \S 4.7). The loss in this case can be written as a functional over the space of measures of the temporal trajectory of the neural preactivations. As in the deep feedforward network case, this loss functional is generally not convex. The KKT condition for a recurrent network depends on the full temporal trajectory of preactivations (SI \S 4.7), rather than factorizing cleanly at each layer as in the feedforward case (SI \S 4.6). Nevertheless, we find that for whitened inputs and the data model in Section \ref{subsec:Data-and-network}, the learned representation at the last timestep remains abstract (Fig.~\ref{fig:6}E). 

\section{Discussion}
The low-dimensional disentangled abstract representations of task-relevant variables have been observed in multiple brain areas of multiple species (Fig.\ref{fig:1}, \cite{bernardi2020geometry,tang2023geometric,fascianelli2024neural,o2023representational,mishchanchuk2024hidden}). Despite the ubiquity of this type of representational geometry, we still do not know the network mechanisms that lead to its formation. Here, we showed that abstract representations of the latent variables naturally emerge in simple feedforward neural networks when the networks are optimized on tasks whose output (i.e. behavior readout) depend on those latent variables. In these networks, the geometry of the hidden layer reflects the geometry of the outputs (labels, Fig.~\ref{fig:1}-\ref{fig:1b}). The input geometry or input encoding also has an important effect on the representational geometry learned in the hidden layer (Fig.~\ref{fig:2}C, Fig.~\ref{fig:4}D-F and \cite{alleman2024task,johnston2024modular}): when the input geometry aligns with the abstract output geometry, it is not surprising that the hidden representation will also be abstract (Section \ref{subsec:Target-aligned-input}). It is less obvious that whitened inputs (Section \ref{subsec:Whitened-input-+}-\ref{subsec:Orthogonal-input-+}), despite not being aligned with any specific low-dimensional representation geometry, nonetheless facilitate the emergence of low-dimensional abstract neural representations. One possible intuition is that this facilitation arises from the maximal dimensionality of the whitened data, which allows the hidden-layer neural representation to "move around more freely" in the high-dimensional space and inherit the low-dimensional structure from the output. In the brain, these high-dimensional representations are likely the result of some form of 'recoding' \cite{marr1991simple,marr1991theory}, which is probably implemented in the hippocampus \cite{gluck1998psychobiological,benna2021place,priestley2022signatures,boyle2024tuned,sun2025learning}. Earlier work has shown that dimensionality expansion increases the separability between different stimuli and helps discrimination \cite{mcclelland1996considerations,barak2013sparseness,litwin2017optimal}. Here, our result suggests that these expansion layers offer another benefit: they facilitate better representation learning in downstream brain areas. These learned low-dimensional representations would support sample-efficient generalization for novel tasks \cite{bengio2013representation,tschannen2018recent}. 


In order to investigate the optimal neural representation formed in the network, we developed a mathematical framework that maps the original problem of weight space optimization to an optimization problem over the neural preactivations [Fig.~\ref{fig:1c}A-C, Eq.~\eqref{eq:loss_h}]. The new problem corresponds to an effective model where neurons interact with each other through a "mean-field" induced by the overall distribution of neural activity [Eq.~\eqref{eq:loss_rho}, Fig.~\ref{fig:1c}]. We derive the KKT condition for the optimal distribution of neural activity [Eq.~\eqref{eq:KKT}]. For a large class of input kernels, the KKT condition is equivalent to a nonnegative quadratic programming problem, which we solve in closed form when the output geometry is abstract (SI \S 2.2). Crucially, because the new optimization problem is convex, any solution of the KKT condition is automatically a global minimum of the loss function. We show that, in many cases, all such solutions correspond to an abstract neural representation. To our knowledge, this provides the first theoretical result showing the robust emergence of abstract representation in nonlinear neural networks, when trained to binary decision tasks closely related to prior neuroscience experiments (Section \ref{subsec:Data-and-network}, \cite{bernardi2020geometry,courellis2024abstract}). These results, together with the analytical framework developed to derive them, constitute the main innovations of this work.

Our work provides insights into why abstract representations appear in the brain \cite{bernardi2020geometry,tang2023geometric,fascianelli2024neural,o2023representational,mishchanchuk2024hidden}. More broadly, our analytical framework provides a general tool for characterizing optimal representations across a wide range of tasks studied in the machine learning literature. It offers a complementary perspective to existing approaches for analyzing learning in two-layer neural networks. Below, we compare our approach with two major streams of related work (see a more comprehensive discussion in SI \S 5):

A substantial body of prior work has investigated two-layer networks from the perspective of Bayesian neural networks. Leveraging tools from statistical physics, these studies have shown that certain scaling limits of the model lead to Gaussian equivalence properties \cite{goldt2020modeling,li2021statistical,pacelli2023statistical}, wherein the neural preactivations follow a joint Gaussian distribution, a regime closely linked to the kernel or lazy regime \cite{pmlr-v125-woodworth20a,farrell2023lazy,jacot2018neural}. More recent work has explored alternative scaling limits \cite{yang2020tensor,yang2021tensor,van2025coding,lauditi2025adaptive} that are connected to the mean-field limit of deep networks. While these results primarily concern the (infinite-width or infinite-input-dimension) scaling limits, our framework enables direct analysis of optimal solutions in finite-width networks with finite-dimensional inputs. The results for finite-width networks allow us to derive results on the optimal representation in various infinite-width limits (SI \S 5.1). Furthermore, whereas Bayesian approaches typically average the network properties over the posterior distribution, our method provides insights into the structure of individual global minima of the loss function. We comment that when deriving the scaling limit from our finite-width results, we first take the infinite temperature limit $\beta \rightarrow +\infty$ followed by infinite-width limit $M\rightarrow +\infty$. As a consequence, the resulting network always learns the low-dimensional features in the data. This is different from the usual Bayesian network setting where $M\rightarrow +\infty$ is taken first and then $\beta \rightarrow +\infty$, where the feature learning only happens when using a special weight scaling in the loss function.

Another series of work examined the learning dynamics of these networks. Earlier works focused on perceptron \cite{seung1992statistical,engel2001statistical,patel2025rl}, deep linear networks \cite{saxe2013exact,saxe2019mathematical} and narrow two-layer network in the teacher-student setup \cite{saad1995exact,saad1996learning,saad1995line}. Recently, the learning dynamics of overparametrized two-layer nonlinear networks were studied both in the mean-field regime \cite{mei2018mean,chizat2018global,rotskoff2018neural,sirignano2020mean,bordelon2022self} and in finite-width settings \cite{boursier2022gradient}. These recent analyses on nonlinear networks are typically only tractable for a one-dimensional output. However, abstract representations (Fig.~\ref{fig:1}) concern the relationship between the hidden representations of different output dimensions (Section~\ref{subsec:Parallelism-score-measures}). So we adopt a different approach here: rather than tracking the entire training trajectory, we directly analyze global minima of the loss function [Eq.~\eqref{eq:TwoLayerNN_loss}]. An intriguing future direction would be to extend the existing results for learning dynamics with one-dimensional outputs, to those tasks in Section~\ref{subsec:Data-and-network} that require multi-dimensional outputs, and investigate how the hidden representation kernel evolves during training.

A byproduct of our analysis is a characterization of single-neuron selectivity in the optimal network solution. This is determined by the set of optimal preactivation vectors $A(K_X,K_Y)$ [Eq.\eqref{eq:h_opt}, \eqref{eq:opt_hidden_kernel_multi} and \eqref{eq:opt_preactivation_2nd_class}], whose components specify the neural response to each stimulus in the training set. A longstanding question in neuroscience is whether neurons exhibit “interpretable” tuning—activity explained by a single task-relevant variable, or “mixed” tuning, where responses depend on combinations of multiple variables \cite{fusi2016neurons}. Experimental data and models have suggested that the tuning type depends on details of task structure \cite{johnston2024modular,whittington2022disentanglement,dorrell2024range,dorrell2026convex}, brain regions \cite{posani2025rarely}, and species. For the task and network architecture studied here (Section \ref{subsec:Data-and-network}), we find that the nonlinearity in the hidden layer plays a key role (see also \cite{alleman2024task}): depending on its form, the task optimization yields either distinct neural modules tuned to individual binary latent variables or neurons with mixed selectivity to multiple latent variables. In the brain, the neuron nonlinearity varies across regions due to differences in single-neuron biophysical and morphological properties \cite{rauch2003neocortical}, which may explain the heterogeneous single-neuron selectivity observed in experimental data.

Despite such variability on the single-neuron selectivity, we demonstrated that in the large-network limit, the emergence of abstract representations on the population level is robust to the specific form of single-neuron nonlinearities. This universality result offers a potential explanation for recent experimental observations of abstract representations across diverse brain areas. Previous studies have shown that neural network models, under certain scaling limits, exhibit Gaussian universality or Gaussian equivalence properties, typically attributed to central limit theorem-like mechanisms \cite{HuUniversity2022,goldt2020modeling}. In contrast, the optimal preactivation distributions ($\rho_*$) in our model are non-Gaussian, suggesting a different origin of universality. We believe that the shared task structure [Section \ref{subsec:Data-and-network}] on which all these models are trained is the underlying reason for this universal abstract representation. This insight aligns with recent empirical findings that networks with different architectures, when trained on similar or related tasks, often converge to similar neural representations, a phenomenon referred to as the Platonic representation hypothesis \cite{li2015convergent,kornblith2019similarity,huh2024position}. Our results thus provide a tractable mathematical model for investigating this hypothesis and show that neural representations for a task on the population level (in terms of representation kernel), can be relatively insensitive to single-neuron-level response details. 


Finally, the order parameter $\rho$, defined as the distribution over neural preactivations, provides a powerful tool for exploiting the permutation symmetry inherent in these network models \cite{bach2017breaking,mei2018mean,sirignano2020mean,chizat2018global}. We anticipate that this formulation will not only aid in analyzing the optimal solutions in the networks studied here but may also be extended to investigate a broader class of permutation-symmetric problems, including ResNet and transformer models in modern AI systems, as well as to analyze learning dynamics in network models trained with biological learning rules.


\section{Acknowledgement}
This work is supported by NSF DBI-2229929 (ARNI), NIH NINDS K99NS138578 (WJJ), the Simons Foundation (542983SPI, SF), the Swartz Foundation, the Gatsby Charitable Foundation, and the Kavli Foundation.

\bibliographystyle{plain}
\addcontentsline{toc}{section}{\refname}
\bibliography{Ref}




\FloatBarrier   
\clearpage

\end{document}